\documentclass[aps,10pt,pra,twocolumn,groupedaddress,showpacs,showkeys,amsfonts]{revtex4-1}
\usepackage{amsfonts}
\usepackage{amsmath}
\usepackage{amssymb}
\usepackage{graphics,graphicx}
\usepackage{epsf}
\usepackage{subfigure}
\usepackage{hyperref}
\usepackage[all]{hypcap}
\usepackage{color}
\newcommand{\eq}[1]{Eq.~\eqref{#1}}
\newcommand{\eqs}[1]{Eqs.~\eqref{#1}}
\newcommand{\seq}[1]{Sec.~\ref{#1}}
\newcommand{\app}[1]{App.~\ref{#1}}
\newcommand{\fig}[1]{Fig.~\ref{#1}}
\newcommand{\figs}[1]{Figs.~\ref{#1}}
\newcommand{\be}{\begin{equation}}
\newcommand{\ee}{\end{equation}}
\newcommand{\bem}{\begin{multline}}
\newcommand{\bea}{\begin{align}}
\newcommand{\eea}{\end{align}}

\def\cro#1{\left[ #1 \right]}
\def\pare#1{\left( #1 \right)}

\begin{document}

\title{Phase-space study of surface-electrode Paul traps: Integrable, chaotic, and mixed motions}

\author{V. Roberdel$^1$}
\author{D. Leibfried$^2$}
\author{D. Ullmo$^1$}
\author{H. Landa$^{1,3}$}
\email{haggaila@gmail.com}
\affiliation{$^1$LPTMS, CNRS, Univ.~Paris-Sud, Universit\'e Paris-Saclay, 91405 Orsay, France
\\$^2$National Institute of Standards and Technology, 325 Broadway, Boulder, Colorado 80305, USA \\ $^3$ Institut de Physique Th\'{e}orique, Universit\'{e} Paris-Saclay, CEA, CNRS, 91191 Gif-sur-Yvette, France}

\begin{abstract}
We present a comprehensive phase-space treatment of the motion of charged particles in electrodynamic traps. Focusing on five-wire surface-electrode Paul traps, we study the details of integrable and chaotic motion of a single ion. We introduce appropriate phase-space measures and give a universal characterization of the trap effectiveness as a function of the parameters. We rigorously derive the commonly used (time-independent) pseudopotential approximation, quantify its regime of validity and analyze the mechanism of its breakdown  within the time-dependent potential. The phase space approach that we develop gives a general framework for describing ion dynamics in a broad variety of surface Paul traps. To probe this framework experimentally, we propose and analyze, using numerical simulations, an experiment that can be realized with an existing four-wire trap. We  predict a robust experimental signature of the existence of trapping pockets within a mixed regular and chaotic phase-space structure. Intricately rich escape dynamics suggest that surface traps give access to exploring microscopic Hamiltonian transport phenomena in phase space.

\end{abstract}


\maketitle

\section{Introduction and main results}\label{Sec:MainResults}

Surface electrode Paul traps are becoming increasingly widespread, in large part due to their potential for scalable microfabrication \cite{1367-2630-12-3-033031,1367-2630-13-7-075018}. Introduced only recently \cite{Chiaverini:2005:SAI:2011670.2011671,PhysRevLett.96.253003}, they differ from traditional linear and hyperbolic Paul traps by the strong nonlinearity of the potential, leading to complex and partly chaotic motion even for a single ion. Furthermore, the small scale of surface traps implies a limited trapping volume, and the importance of perturbations, intrinsic or extrinsic to the trap, grows significantly. Therefore, on the practical side, a comprehensive understanding of the different dynamical regimes in these traps is important for optimizing ion loading, cooling, storage and manipulation.
From a more fundamental perspective, surface traps offer a unique setup for studying the interplay of nonlinearity, chaos, and microscopic stochastic forces.

Nonlinear and stochastic ion dynamics have attracted fundamental and practical interest since the first crystallization of charged particles in electrodynamic fields \cite{FirstCrystal}.
The first systematic studies of chaos in Paul traps followed the demonstrations of a cloud-crystal phase transition with small clusters \cite{Walther1987,PhysRevLett.59.2935} and clouds  \cite{Walther1988}. Since Paul traps are based on (periodic) time-dependent electric fields, the system does not conserve energy and so particles can eventually heat up and escape the trap. Deterministic chaos was found to be a predominant source of heating that balances the laser cooling and stabilizes the cloud phase. The smallest system with {appreciable} nonlinearity (and hence, possibly, chaotic motion) in the earlier Paul traps was that of two ions, and it was studied in detail \cite{Brewer1988,Brewer1990,Brewer1994,Brewer1995,
Frequency_Locked_Orbits1993,Frequency_Locked_Orbits1994,riddled_basins}, within a time-independent approximation \cite{Blumel1989,Blumel_rf_resonances}, including an analysis of all integrable cases \cite{Baumann1992,BaumannComment1,BaumannComment2,BaumannReply}.
Due to their importance in high-accuracy quantum applications, the effects of the periodic driving (known as the `micromotion') are extensively studied in various regimes \cite{berkeland1998minimization,PhysRevLett.81.3631, rfions,zigzagexperiment, Landa2014, arnold2015prospects, keller2015precise,keller2017optical,keller2018controlling}. {The nonlinear (and time-dependent) regimes of motion in a surface trap, remain however largely unexplored.}

In this paper we develop a phase space framework for studying {classical} dynamics of charged particles in surface Paul traps. We introduce appropriate phase space measures that allow us to quantify and compare the trap's effectiveness at different parameter values. Our quantitative results apply directly to any 5-wire trap of any size (for the model that we consider). The general conclusions of the study apply to a broad range of surface traps and present a systematic approach to the investigation of the nonlinear and time-dependent motion. 

 We consider an ion trapped in vacuum above a set of planar electrodes in the `5-wire' configuration, carrying a combination of direct current (DC) and radio-frequency (rf) periodically modulated voltages (see \seq{sec:5-wire surface trap} and \fig{fig:trap}). The axial $x$ motion (parallel to the electrodes' symmetry axis) is that of a decoupled, time independent harmonic oscillator with frequency $\omega_x$, the axial harmonic trapping frequency. The $yz$ motion is coupled, nonlinear, and driven periodically in time, and hence the phase space $\left\{y,z,p_y,p_z\right\}$ whose dimension is 4 (with $p_y$ and $p_z$ the canonical momenta), has to be in practice increased  to include the time $t$, making it effectively five-dimensional (5D).

5D phase space is a big place \footnote{In paraphrase on the quote by Carlton Cave on Hilbert spaces, the complete characterization of the motion in 5D phase space can be very challenging}.
Moreover, the number of independent parameters of the full model can become quite high \cite{SurfaceTrapLucas}. However, we show that an essential understanding of the dynamics can be obtained by a systematic study starting from a 2D time-independent reduction, and continuing through more elaborate 3D (which is the effective dimension of a 2D phase space with a time-dependent Hamiltonian) and 4D phase space analysis. This procedure, which we develop in this work, requires also that we limit the parameters at each step.
Hence as a starting point we assume a trap configuration with equal $y$ and $z$ harmonic DC anti-trapping coming from the axial electrode, and no further harmonic DC terms. In this case, an ion starting with the initial conditions $\dot{y}=y=0$ (in the coordinate parallel to the electrodes' plane), remains at $y=0$, and the $z$ direction (normal to the electrode surface) decouples exactly. We study this motion in detail in \seq{Sec:Hamiltonian1D}, also assuming at first that there is no further `bias' DC voltage on the rf electrodes. Then the dynamics depend on two parameters -- the rf voltage $U_{\rm{rf}}$, and the axial DC trapping voltage, $U_{\rm{DC}}\propto\omega_x^2$.
Studying the rf driven $z$ motion begins with the commonly used approximation that replaces the time-dependent potential with an effective, time-independent `pseudopotential'. The latter depends in this setup on a single nondimensional parameter that we name `the pseudopotential parameter', 
\be \lambda=\left(m\omega_x\Omega w^2/\sqrt{2}eU_{\rm rf}\right)^2\propto U_{\rm{DC}}/(U_{\rm rf})^2,\label{eq:Lambda0}\ee
 where $m$ and $e$ are the ion's mass and charge, $\Omega$ the rf frequency and $w$ the electrodes' width and separation (see \seq{sec:5-wire surface trap} for details). Since the pseudopotential is time-independent, the phase space is 2D, and the motion is integrable without any chaos. Within the pseudopotential approximation, a maximization of the trap depth is misleadingly simple -- it is a monotonously decreasing function of $\lambda$, so it decreases with $\omega_x$ and increases with $U_{\rm rf}$. 

This simple dependence is completely altered within the time-dependent potential. The latter can be thought of as composed of the pseudopotential (characterized by $\lambda$), and a time-dependent perturbation scaled by $U_{\rm rf}$. Since the equations of motion depend explicitly on time, the phase space is now effectively 3D. With $U_{\rm rf}$ below a threshold value, the entire 3D phase space is very close to regular and the pseudopotential approximation is admissible, for any $\lambda$.
Increasing $U_{\rm{rf}}$ gradually makes the phase space `mixed' -- chaotic motion develops inside bounded strips within the regular phase space, and most notably within a large connected `chaotic sea' that reaches from the top of the trap and penetrates the phase space toward the trap center. Once inside the chaotic sea, the ion will quickly escape past the trap's barrier, and therefore increasing $U_{\rm{rf}}$ beyond an optimal value leads to a degradation of the trap's effectiveness. 
We find that the maximal trapping can be obtained by taking $U_{\rm rf}$ at the threshold of validity of the pseudopotential approximation, together with $\lambda\to 0$ (requiring the reduction of the axial trapping $\omega_x$). If $\omega_x$ is constrained to a certain value or range of values, optimal trapping can be obtained by tuning $U_{\rm rf}$ to a $\lambda$-dependent value.

With this  characterization of the 3D phase space of $z$ motion, we turn in \seq{sec:Hamiltonian2D} to the coupled $yz$ motion, starting with the 4D phase space of the pseudopotential approximation. Remarkably, we find that the motion is very close to being completely integrable, i.e.~with almost no chaotic signatures. This is quite surprising since the pseudopotential is highly nonlinear and a-priori there is no reason to expect its integrability. 
We continue by first studying in detail the effect of a bias DC voltage ($U_{\rm{b}}$) applied to the rf electrodes, still within the pseudopotential's 4D phase space. The bias voltage leads first to an increase of the trap depth, but beyond a threshold magnitude causes the gradual destruction of the regular phase space. As a function of the energy, the phase space may split into a few (nearly regular) trapping `islands' separated by chaotic regions. 

Reintroducing the time-dependence of the trap we study how the integrability-breaking perturbation that scales with $U_{\rm rf}$ leads the ion to escape from the trap, due to the combined effect of the bias and time dependence. In the 5D phase space, the regions of regular and chaotic motion can be intertwined in a complicated way. This introduces new possibilities for transport mechanisms in phase space, as we briefly discuss below. At the same time, it presents a challenge for visualizing and exhaustively analyzing the motion. For a practical, first step study of the ion's escape from the trap,  we present a simple measure of phase space volume that enables quantification of these competing processes and their dependence on the three parameters $\lambda$, $U_{\rm rf}$, and $U_{\rm{b}}$. 

We find (in \seq{Sec:rf2D}) that the border of validity of the pseudopotential (for approximately regular motion) found for the $z$ motion,  holds also for the coupled $yz$ motion, restricted to intermediate values of $\lambda$ and $U_{\rm b}$. Above this threshold, all chaotic trajectories escape, and the regular region gradually shrinks as well. 
We also find an optimal regime of the parameters maximizing the (mostly regular) trapping phase space volume of $yz$ motion in the model 5-wire trap. We find that this regime is obtained for $U_{\rm rf}$ just below the threshold of validity of the pseudopotential, with $\lambda \to 0$ (requiring to reduce the axial trapping $\omega_x$), and an optimal intermediate value of $U_{\rm{b}}$. These parameters are distinctly different from a na\"ive pseudopotential approach that would lead to increasing $U_{\rm{b}}$ and $U_{\rm rf}$ as much as possible.

The near-integrability of the pseudopotential for the symmetric, unbiased 5-wire configuration, makes it a natural starting point to analyze dynamics in surface traps, since the motion remains close to integrable for any perturbation that is not too large. This includes the full rf driven motion in 5D phase space, and indeed we confirm that this perturbation can be considered as `small' within the 5D phase space, in the same regime where it is small in the 3D phase space of $z$ motion. A systematic treatment based on the integrable pseudopotential limit of the symmetric 5-wire trap, gives a framework for studying further coupling or asymmetry, or the application to different geometries. The consideration of circular traps  \cite{tabakov2015assembling,li2017realization}, and 
even of electrons guided on a chip \cite{hoffrogge2011microwave}, could be another interesting direction, immediately applicable (the latter is essentially a surface trap, operating at much higher frequencies since the electrons are much lighter). 

Motivated by practical questions on the motion of trapped ions, this work is focused mostly on the ion's escape from the trap. However, from a broader point of view, one can study how transport between different parts of the phase space proceeds within the Hamiltonian motion. As described in the following, for a phase space which is effectively 3D or 4D, the motion is amenable to classification using 2D Poincar\'e sections, and the transport properties can be scrutinized. With more degrees of freedom, there are richer options for transport and there is a fundamental interest in unveiling the involved mechanisms. The ion trap offers a system with excellent prospects for experimenting with this physics, at the level of elementary particles. The trap parameters are highly controllable, the dimensionality of the effective phase space can be  altered, and as we find in the following, further control parameters can be devised. There are some limitations with respect to directly measuring microscopic properties of the motion, that nonetheless can be circumvented.
 
In \seq{sec:experiment} we propose an experiment probing the
nonlinear and chaotic motion of an ion in a surface trap. Our analysis
is based on a 4-wire trap in use at  NIST \cite{wilson2014tunable},
with an electrode configuration that is  more complicated than the model of
the 5-wire trap discussed above. We consider adding a controlled
`tickle' perturbation (a small-amplitude voltage modulation) on one of
the electrodes, at a frequency above resonance with the secular
motion. It gives a robust experimental procedure resulting in a clear
signal with a non-monotonous and sensitive dependence on the
parameters.

 Running numerical simulations of this trap configuration
with real experiment parameters, we find clear evidence (within the
limitations of an initial study), of trapping `pockets' induced in the
trap's phase space. 
We predict a variety of distinct functional forms for the ion's escape
probability from these pockets, which present a challenge for further
theory and experiments on Hamiltonian transport in a mixed phase space
\cite{bohigas1993,PhysRevE.87.012918,PhysRevE.96.032204}. We also find
quantitatively very similar escape probabilities with the
pseudopotential approximation for this trap, indicating that it is
being operated not far from its integrability limit.  
 
The theory presented in this work gives a general framework that can be applied to various surface-electrode traps. With an understanding of the mechanism that
determines the nature of the dynamics, the scope of the  conclusions that we draw [\seq{Sec:Outlook}] extends beyond the particular models here studied. In the Appendices we collect a set of phase space tools that allow one to treat
arbitrary trap potentials which are nonlinear and periodically  driven. 
In particular, we rigorously derive the pseudopotential approximation
for a generic periodic potential, as a canonical transformation from
the original coordinates, that results in a time-independent
pseudo-Hamiltonian (which may a-priori depend on the canonical momenta
\cite{rahav2003time,rahav2003effective}). The derivation of the
pseudopotential from a canonical transformation allows us to keep the
Hamiltonian phase space structure, and in particular the structure of
the invariant tori of the motion (in the near integrable regime). As
briefly discussed in the outlook of \seq{Sec:Outlook}, this Hamiltonian structure
turns out to be invaluable in a treatment of stochastic motion in
phase space, using a Fokker-Planck approach \cite{rfcooling}.

\section{The 5-wire surface trap}\label{sec:5-wire surface trap}

\begin{figure}
\includegraphics[width=3.3in]{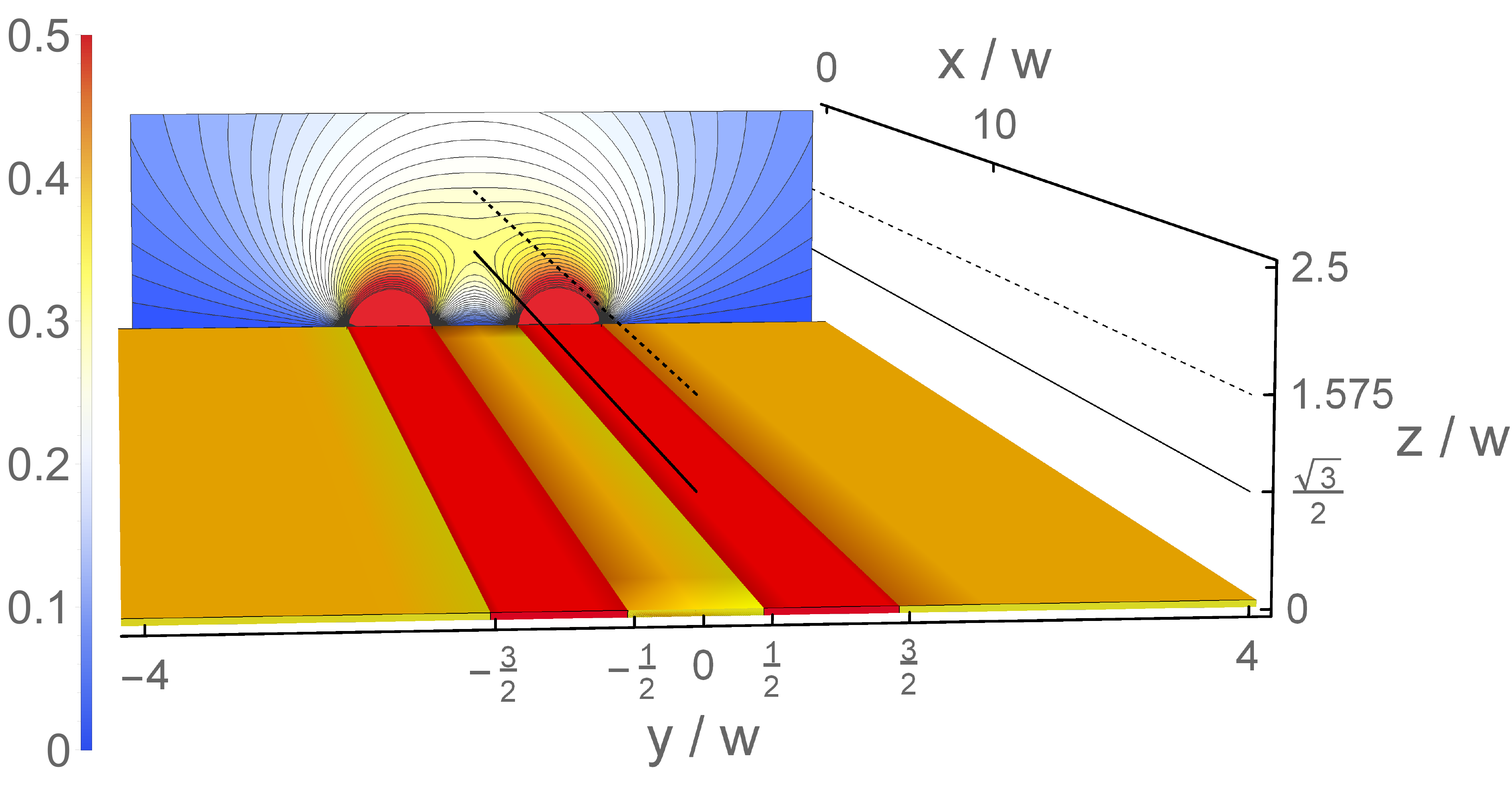}
\caption{Layout and rf-potential of the 5-wire surface trap. All electrodes lie in the $z=0$ plane, with the two electrodes connected to the rf-drive shown in red. They are of width $w$ in the $y$-direction with their center offset by $\pm w$ from $y=0$. Along the $x$-direction they are approximated as having an infinite extent. The remainder of the $z=0$ plane is filled by grounded surfaces, shown in  gold. Setting the rf-electrodes to $1\,$V produces the  potential $V_{\rm 5w}(y,z)$ of \eq{eq:5wnew}, with equipotential lines in the plane $x=0$ shown as a contour plot in the back of the figure. The bar legend shows the potential in units of V. The thick solid black line shows the potential minimum line  and the thick dashed line shows the saddle-line of the potential [with $z=z_s$ and $z=z_u$, respectively, defined in \eq{eq:z_cz_s}]. Not shown in the figure are the electrodes giving rise to the harmonic DC potential of \eq{Eq:Vh}. 
}
\label{fig:trap}
\end{figure}

 In this section we present the details of the trap model and set the notation for the rest of this work. The total trap potential is the sum of two parts; the time dependent potential produced by two rf surface electrodes, and a time independent harmonic potential contribution (quadratic in the coordinates).
As shown in \fig{fig:trap}, the surface electrodes lie in the $z=0$ plane, and are assumed to have an infinite extent along the $x$-axis.
The $y$-axis is perpendicular to the infinite direction of the electrodes in the electrode plane, and the $z$-axis is orthogonal to the electrode plane. The electric potential from the surface electrodes is given by \cite{house2008,wesenberg2008,SurfaceTrapLucas,schmied2011quantum},
\be V_{\rm 5w}(y,z)=\cro{V_{\rm{w}}(y,z;w)-V_{\rm{w}}(y,z;-w)},\label{eq:5wnew}\ee
with the potential due to one planar electrode of width $w$ centered at $+w$ and held at  unit voltage being 
\bem
V_{\rm{w}}(y,z;w)=\\\frac{1}{\pi}\cro{\arctan\pare{\frac{y-w/2}{z}}-\arctan\pare{\frac{y-3w/2}{z}}}.
\label{Vwiredef}\end{multline} 

The field due to the potential of \eq{eq:5wnew} vanishes at a saddle point $\vec{R}_{\rm{0}}=\left\{0,0,z_{\rm{0}}\right\}$ at a height $z_0= \sqrt{3}w/2$ above the electrodes. Since this potential does not trap along the $x$-axis, additional (DC) electrodes must be included. Here we approximate the potential confining the ion along $x$ by an ideal static quadrupole (harmonic) potential along $x$, that leads to anti-trapping in the radial directions.  The total dimensional potential energy of an ion in the trap is 
\bem
\tilde{V}_{\rm{trap}}=e\sum_\alpha \frac{ U_{\rm DC}}{2c_{\alpha}}{\pare{{R}_{\alpha}-{R}_{\rm{0},\alpha}}}^2 \\+e\left(U_{\rm b}-U_{\rm{rf}} \cos\Omega t\right) V_{\rm 5w}(y,z), 
\label{V5wdim}
\end{multline}
where  $\alpha \in \{x,y,z\}$, $\vec{R}$ is the vector coordinate of
the ion, $\Omega$ is the rf frequency, and $e$ the charge of the ion.
$U_{\rm{rf}}$ and $U_{\rm b}$ are the rf and the DC `bias' voltages
(respectively) applied to the rf electrodes, and $U_{\rm{DC}}$
 characterizes the strength of the static harmonic potential. The
geometric properties of the latter are defined by
$c_{\alpha}$ (which are of squared-length dimension), and by the
origin of the static quadrupole that we choose to coincide with the
saddle point of $V_{\rm 5w}$, that is
$\vec{R}_{\rm{0},\alpha}=\left\{0,0,z_{\rm{0}}\right\}$.

Measuring distances in units of the width $w$, and rescaling time by half the rf frequency, $\Omega/2$,
\begin{equation}
 \vec{R}_{\alpha}\rightarrow \vec{R}_{\alpha}/w \qquad t \rightarrow \Omega t /2,
\label{eq:coordandtimerescale}
\end{equation} 
 we find the non-dimensional potential
\be
V_{\rm{trap}}=V_{\rm h}+\left(a_5 -2q_5 \cos 2t\right) V_{\rm 5w}(y,z),
\label{V5wnondim}
\ee
where
\be V_{\rm h}=\frac{1}{2}\left[a_x x^2+a_y y^2+a_z\pare{z-z_{\rm{0}}}^2\right],\label{Eq:Vh}\ee
and the non-dimensional parameters are
\begin{equation}
a_\alpha=\frac{4  e U_{\rm{DC}}}{m  \Omega^2 {c_{\alpha}}}, \quad q_5=\frac{2eU_{\rm{rf}}}{m w^2 \Omega^2}, \quad a_5=\frac{4eU_{\rm{b}}}{m w^2 \Omega^2}.
\label{aq5def}
\end{equation}

 In this setup, the motion in the $x$ direction is that of a simple harmonic oscillator (with frequency $\omega_x$), decoupled from the motion in the radial $yz$ plane. As a consequence of the Laplace equation, the harmonic axial DC voltage gives rise to radial anti-trapping, which we take to be symmetric;
\begin{equation}
a_z=a_y=-\frac{1}{2}a_x,\qquad a_x > 0.
\label{axayaz}
\end{equation} 

The linearized secular frequencies of the ion near the center of the trap are given by \cite{leibfried2003}
\begin{equation}
\omega_{\alpha}= \nu_\alpha \frac{\Omega}{2},
\end{equation}
where $\nu_\alpha\left(a_\alpha,q_\alpha\right)$ is the non-dimensional characteristic exponent of the corresponding Mathieu equation, with parameters $a_\alpha,q_\alpha$. The latter ($q_\alpha$) can be obtained by a linearization of the trap potential $V_{\rm trap}$ of \eq{V5wnondim} around $\{y=0,z=z_0\}$, giving, for $a_5=0$,
\be
q_x=0,\qquad q_z=-q_y=\frac{2}{\sqrt{3}\pi}q_5.
\ee
In the limit $a_\alpha,q_\alpha^2 \ll 1$ we can approximate
\begin{equation}
\nu_{\alpha}\approx\sqrt{a_\alpha + q_{\alpha}^2/2},
\label{eq:omegas}
\end{equation}
which results in
\begin{equation}
\omega_{x}=\sqrt{a_x}\frac{\Omega}{2}, \qquad \omega_z \approx \sqrt{- a_x + q_{z}^2}\frac{\Omega}{2\sqrt{2}}.
\label{eq:omegax}
\end{equation}

For the figures and numerical analysis in this paper, we consider both non-dimensional parameters and physical parameters that give a specific example of real-world values. All the dimensional values are given for $^9{\rm Be}^+$ ions in a trap with electrode width and rf frequency (that appear in \eq{eq:coordandtimerescale}), given by
\be w=50\,{\rm \mu m},\qquad \Omega=2\pi\times 100\,{\rm MHz}.\label{eq:physicalparams}\ee

\section{Hamiltonian motion in one spatial dimension}\label{Sec:Hamiltonian1D}

In this section, we consider the unbiased configuration $U_{\rm b}=0$
(i.e.\ $a_5 =0$), and characterize the motion of the ion in the rf
potential, and the corresponding pseudopotential, of the $z$ spatial
coordinate  orthogonal to the electrode plane. In
\seq{subsec:reduction} we introduce the notation, and in
\seq{subsec:Phase-space visualisation} we present the possible types
of phase space structures and discuss the qualitative
implications. Then in \seq{sec:psqevolution} we define quantitative
measures that enable characterization of the trap's dependence on its
parameters. The effect of the bias voltage
$U_{\rm b}$ will be studied in \seq{sec:Hamiltonian2D}, for the
coupled $yz$ motion.

\subsection{Reduction of the equation of motion}\label{subsec:reduction}


Setting $y=\dot{y}=0$ in the equations of motion derived from
\eq{V5wnondim} results in $y(t)=0$ for all times, and thus the $z$
motion can be studied independently. Since $a_5=0$, the potential depends on two
parameters, $a_x\propto\omega_x^2$ and $q_5$, that are proportional to
the static quadrupole and the rf voltages respectively [\eq{aq5def}].  
With $y=0$ in \eq{V5wnondim}, the rf potential can be simplified to   
\bem
V_{\rm{rf}}^{\rm 1D}=a_z\frac{{\pare{z-z_{\rm{0}}}}^2}{2} \\-\frac{4}{\pi}q_5\cos 2 t \cro{\arctan(\frac{-1/2}{z})-\arctan(\frac{-3/2}{z})}.
\label{1Ddimrfpot} 
\end{multline}
In the rest of the paper we refer to this potential as the 1D `rf potential'.

Since the rf frequency sets the fastest frequency scale, the method of averaging is a natural approach allowing one to simplify the description of the motion, and reduce the equations of motion to an autonomous system. Using this method it is possible to derive an effective time-independent pseudopotential approximation to the rf potential, that is well known \cite{PhysRev.170.91,amini2008}. In \app{app:pseudopot} we present a derivation of the pseudopotential as a time dependent canonical transformation, from the original phase space coordinates of motion within the rf potential, to new coordinates. This transformation eliminates the $\pi$-periodic component of the position and momentum, while preserving the Hamiltonian structure (with a new time-independent Hamiltonian).{ Thus, although the  new Hamiltonian is time independent, the new variables are actually time dependent functions of the original ones (and vice versa).} In the following subsections we will consider in detail the regime where the pseudopotential gives a simpler and accurate approximation, and its limits of applicability.

 The 1D pseudopotential approximation to \eq{1Ddimrfpot}, derived in \app{app:pseudopot}, is 
\begin{equation}
{V}_{\rm{pseudo}}^{1{\rm D}}=a_z \frac{(z-z_{\rm{0}})^2}{2}+q_{5}^2\frac{16 \left(3-4z^2\right)^2}{\pi^2\left(9+40z^2+16z^4\right)^2}.
\label{2paramSurfacePseudoPot1D}
\end{equation}
By a second rescaling of time using 
\be t \rightarrow q_5t\label{tq_5rescale},\ee
we get the rescaled pseudopotential, 
\begin{equation}
{V}_{\lambda}^{\rm 1D}\equiv \frac{{V}_{\rm{pseudo}}^{\rm 1D}}{q_5^2}=-\lambda \frac{(z-z_{\rm{0}})^2}{2}+\frac{16 \left(3-4z^2\right)^2}{\pi^2\left(9+40z^2+16z^4\right)^2}
\label{1paramPseudoPot1D}
\end{equation}
governed by a single pseudopotential parameter defined equivalently to \eq{eq:Lambda0} by 
\begin{equation}
\lambda=-\frac{a_z}{ q_{5}^2}=\frac{a_x}{2 q_{5}^2}>0,
\label{definelambda}
\end{equation}
where we have used the fact that in the symmetric case that we consider, $a_z=-a_x/2<0$, as in \eq{axayaz}.

The pseudopotential has two fixed points, defined as the solutions of 
\be{ d }{V}_{\rm{pseudo}}^{\rm 1D}/{ d }z=0.\label{eq:dEdz}\ee
 The first point, $z_s$, is the stable fixed point at the center of the trap (that is, the pseudopotential minimum). The second point, $z_u$, is an unstable fixed point, a local maximum of the pseudopotential, beyond which the ion escapes the trap. As discussed above, we assume a configuration such that the $z$ origin of the harmonic term in \eq{1paramPseudoPot1D} coincides with the point where the force from $V_{\rm 5w}$ is 0. In that case the stable fixed point is independent of the parameters $a_x$ and $q_5$: Within the pseudopotential as well as the rf potential,   $z_{\rm{0}}=z_{s}$ for all $\lambda$. For $\lambda=0$, $z_u$ is maximal, while for $\lambda>0$, $z_u$ moves toward $z_s$. The numerical values of the fixed points are given by
\begin{align}
&z_s=\frac{\sqrt{3}}{2}\approx 0.866,
&&\left.z_u\right|_{\lambda=0}=\sqrt{3/4+\sqrt{3}} \approx 1.575.\label{eq:z_cz_s}
\end{align}

The linearization near $z_s$ of ${V}_{\rm{pseudo}}^{1{\rm D}}$ of \eq{2paramSurfacePseudoPot1D} gives the harmonic oscillator potential
\be {V}_{\rm{pseudo}}^{1{\rm D}} \to {V}_{\rm{h.o.}}^{1{\rm D}}\equiv \frac{1}{2}\nu_z^2 z^2,\label{Eq:ho}\ee
and the linearization of ${V}_{\rm{rf}}^{1{\rm D}}$ of \eq{1Ddimrfpot} gives the Mathieu oscillator potential
\be {V}_{\rm{rf}}^{1{\rm D}} \to {V}_{\rm{M.o.}}^{1{\rm D}}\equiv \frac{1}{2}(a_z-2q_z\cos2t) z^2,\label{Eq:Mo}\ee
with characteristic exponent $\nu_z(a_z,q_z)$ that coincides with $\nu_z$ appearing in \eq{Eq:ho}, within the approximation of \eq{eq:omegas}.

\subsection{Phase space structure}\label{subsec:Phase-space visualisation}

\begin{figure}
\subfigure[\,Pseudopotential]{
\includegraphics[trim = 0cm 0cm 2.5cm 0cm, clip,width=0.22\textwidth]{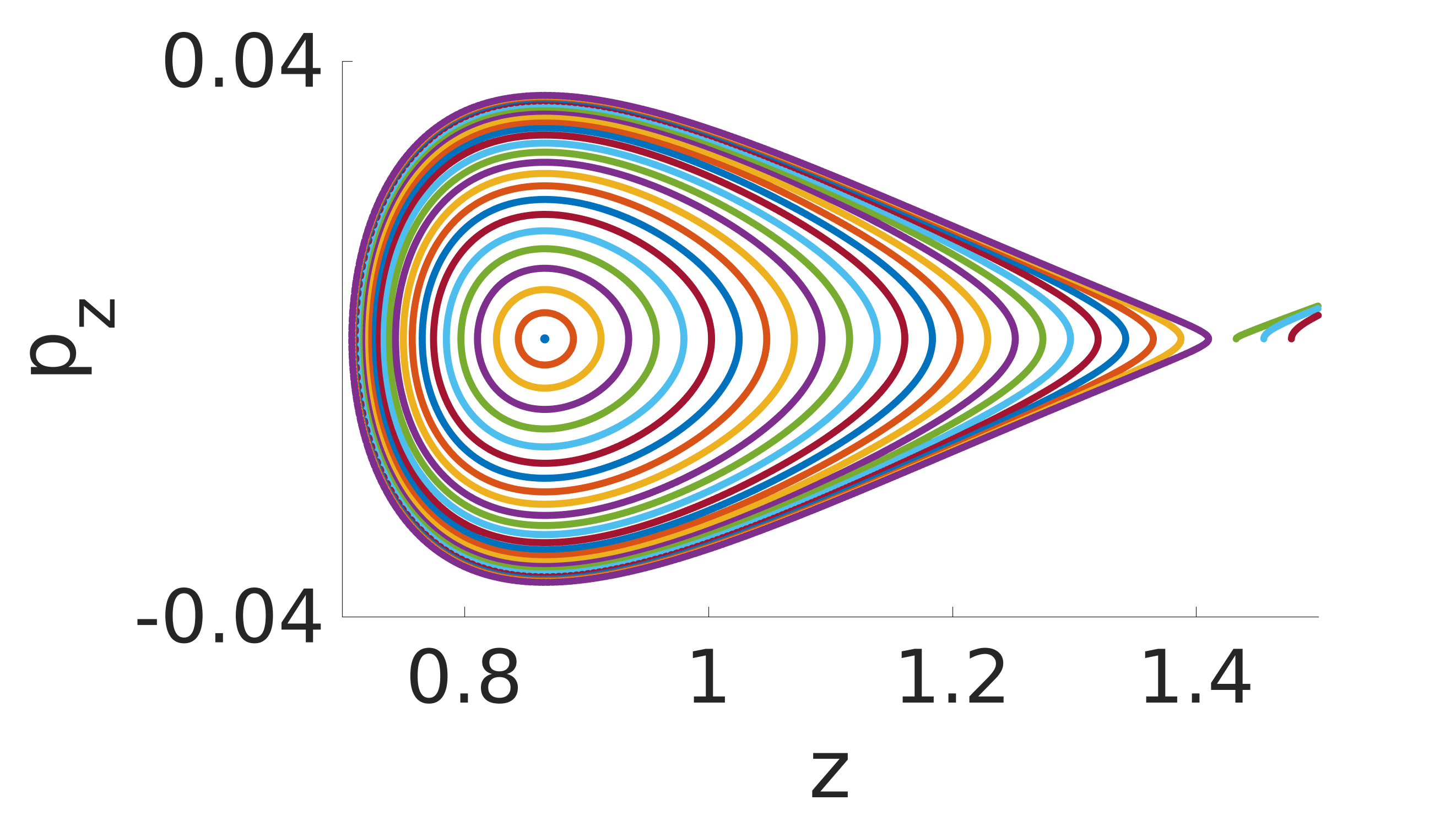}
\label{fig:0-40-P}}
\subfigure[\,Rf potential]{\includegraphics[trim = 0cm 0cm 2.5cm 0cm, clip,width=0.22\textwidth]{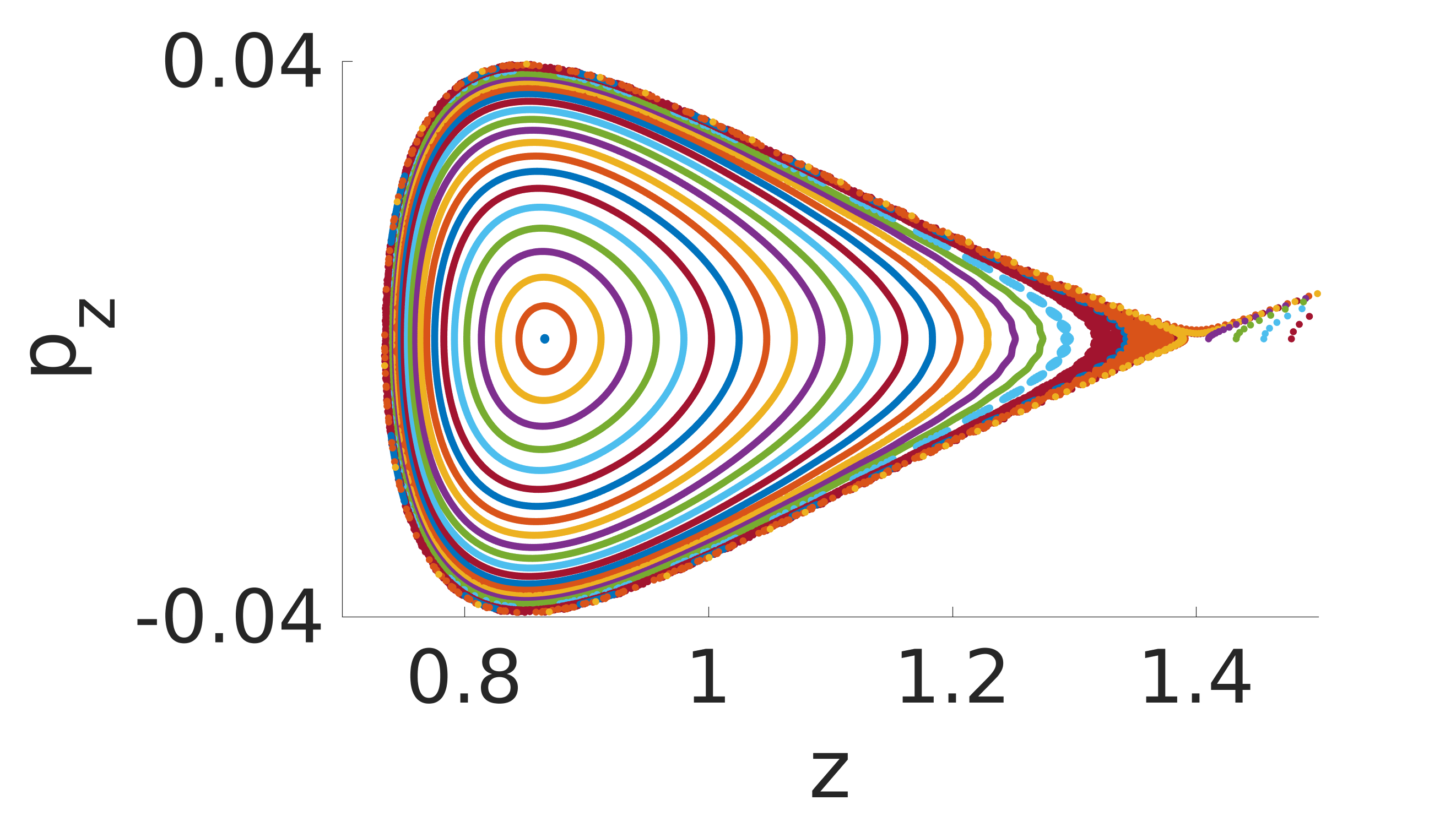}
\label{fig:0.0016-30-Rf}}
\bigskip
\subfigure[\,Rf potential]{\includegraphics[trim = 1.9cm 0cm 2.5cm 0cm, clip,width=0.22\textwidth]{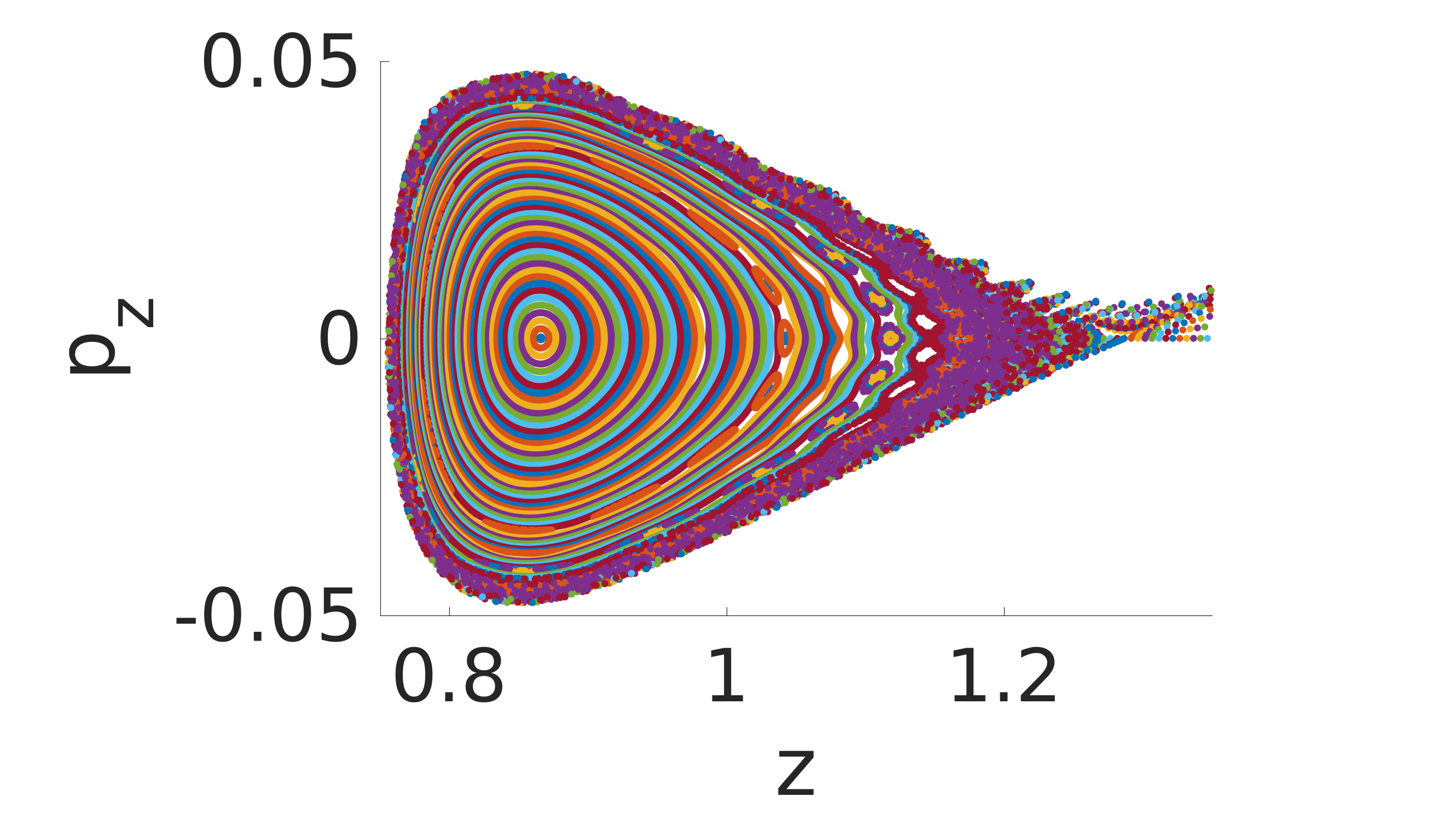}
\label{fig:b1}}
\subfigure[\,Zoom in]{\includegraphics[trim = 1.9cm 0cm 2.5cm 0cm, clip,width=0.22\textwidth]{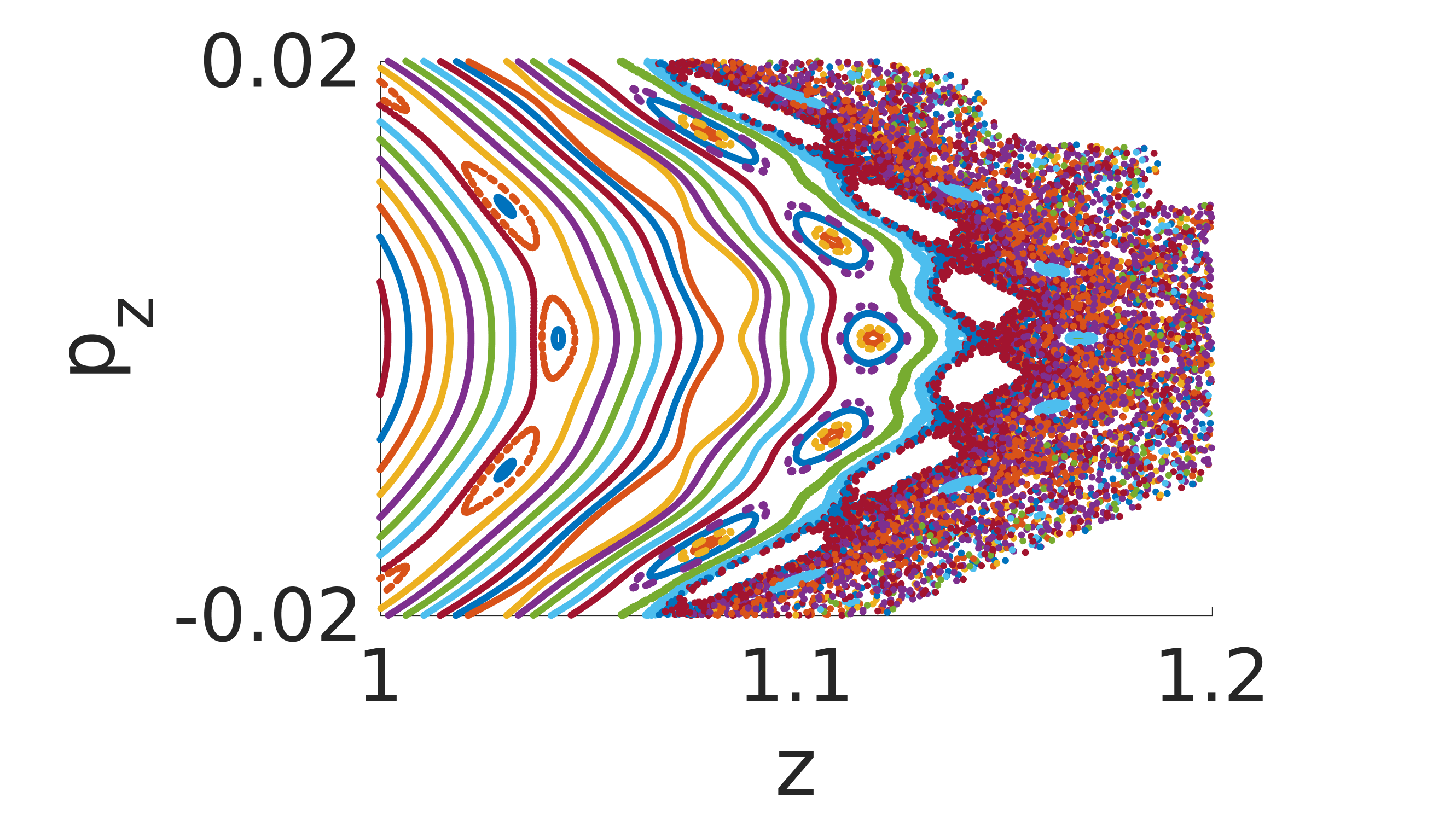}
\label{fig:b2}}
\caption{(a) Phase space trajectories for motion in the non-dimensional pseudopotential of \eq{2paramSurfacePseudoPot1D} with $\surd{\lambda}=0.035$ and $q_5=0.65$ [$\omega_x=2\pi\times 2\,{\rm MHz}$, $\omega_z\approx 2\pi\times 8.4\,{\rm MHz}$, $U_{\rm{rf}}=30$V, see also \eq{eq:physicalparams}]. For each initial condition the ion moves continuously along a curve. The stable ($z_s$) and unstable ($z_u$) fixed points can be identified at the center and at the right tip, together with some escaping trajectories. (b) Stroboscopic map taken at $t = 0 \pmod {\pi}$, of the motion in the rf potential [\eq{1Ddimrfpot}] for the same parameters. The ion returns to this 2D `cut' of its phase-space every $\pi$-period, and for regular motion, all points started with a given initial condition lie on a curve. Small regions of chaotic motion, filling an area in phase-space, can be identified nearby the slightly lowered unstable fixed point. (c) Similarly, for $\surd{\lambda}=0.068$, $q_5=0.868$ [$\omega_x=2\pi\times 4.18\,{\rm MHz}$, $\omega_z\approx 2\pi\times 11.1\,{\rm MHz}$, $U_{\rm{rf}}=40$V], the motion is mostly regular inside the last unbroken torus (with resonance island chains surrounded by very thin chaotic strips, between unbroken tori), beyond which a chaotic `sea' emerges from the region near the unstable fixed point. (d) A zoom into the region around the last unbroken torus of (c).} \label{fig:StrobPlots}
\end{figure}


The pseudopotential is time-independent, and therefore the total energy is conserved in time. Furthermore, conservative motion in one spatial dimension is integrable, and the motion of the ion in the 2D phase space $\left\{z,p_z\right\}$ 
is restricted to a closed curve for bounded motion, see \fig{fig:StrobPlots}(a). Here, $p_z=\dot{z}$ is the canonical momentum, with the mass absorbed into the parameters in \eq{aq5def}.

The rf potential is, in contrast, time-dependent and $\pi$-periodic, so the motion can be thought of as
occurring within a 3D phase-space defined by $\left\{z,p_z,t\right\}$,
with the $t$ dimension periodic. The ion trajectories can be studied
by constructing a 2D `cut' of phase-space taken at fixed value of $t
\pmod {\pi}$, see \fig{fig:StrobPlots}(b). The continuous Hamiltonian
motion as a function of time, is then `stroboscopically' projected to a
discrete map of phase-space onto itself at $\pi$ time-intervals. The
ion  does not follow continuously as a function of time any curve in this
planar surface of section, but rather revisits this plane every
$\pi$-period of the external drive.

\begin{figure}
\includegraphics[width=0.48\textwidth]{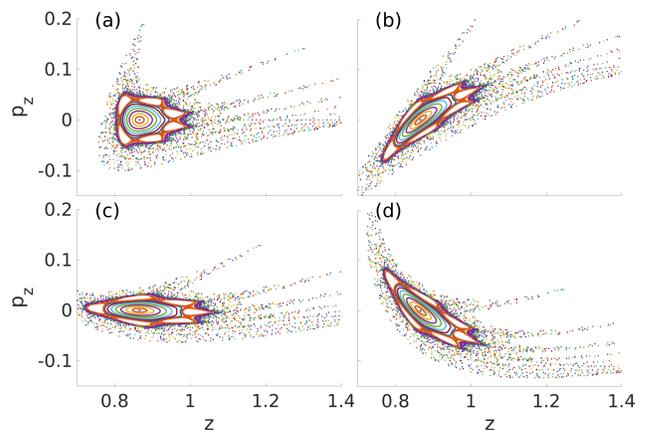}
\caption{Stroboscopic map of the time-dependent phase space for $\surd{\lambda}=0.015$ and $q_5=1.30$ [$\omega_x=2\pi\times 1.41\,{\rm MHz}$, $\omega_z\approx 2\pi\times 17.8\,{\rm MHz}$, and $U_{\rm{rf}}=60$V], at (a) $t = 0 \pmod {\pi}$, (b)  $t \approx 0.26\pi \pmod {\pi}$, (c) $t \approx 0.53\pi \pmod {\pi}$ and (d) $t \approx 0.8\pi \pmod {\pi}$. Although $\lambda$ is small and therefore the unstable fixed point is close to its maximal position, the chaotic region connected to it is extremely large. The distortion of phase space during a micromotion cycle can be clearly seen. The area of every set of points that maps onto itself under the Hamiltonian evolution is invariant.}
\label{fig:0_0008-60-Rf}
\end{figure}

In contrast to the pseudopotential, within the time-dependent
potential two  different types of motion can arise -- regular and
chaotic. For bounded regular motion, the ion's return points to the 2D
surface of section lie on a closed, 1D curve, or a finite set of such
curves in the case of a chain of resonance `islands'. In the latter case, in each
period the ion revisits a different island from the chain in a fixed
order. For chaotic motion, the ion's return points will in the limit
of infinite time fill a region with nonzero area. The structure of
unbroken tori (invariant curves), chaotic strips and resonance island
chains is fractal [repeating itself in smaller and smaller scales --
see \fig{fig:StrobPlots}(c)-(d)]. Each closed curve and every
connected region of chaotic trajectories, are invariant under the time
evolution. Therefore a stroboscopic map taken at any value of
$t\pmod{\pi}$, will contain the corresponding structures with the same
enclosed area, distorted in time with periodicity ${\pi}$, see
\fig{fig:0_0008-60-Rf}. This important property of the time evolution
will be used below. 
 
Panels (a)-(b) of \fig{fig:StrobPlots} show the similarity between the pseudopotential phase space and that of the rf potential, for small $q_5$. This can be understood by decomposing the rf potential as a sum of its time-independent averaged pseudopotential (whose phase space structure depends on the single parameter $\lambda$), and a time-dependent perturbation, whose amplitude is controlled by $q_5$. Therefore we expect the rf potential to be well approximated by the pseudopotential in the limit of small $q_5$, for any $\lambda$. The pseudopotential is integrable, and the measure of chaotic trajectories in the phase space increases with $q_5$. When $q_5$ is small the phase space is composed of mostly regular motion and as it is increased, an increasing area of phase space becomes chaotic. Panels (c)-(d) of \fig{fig:StrobPlots} show the stroboscopic map with a moderate value of both $\lambda$ and $q_5$, for which there is significant deviation from integrability with the aforementioned mixed structure of regular tori, chaotic strips, resonance islands, and a large chaotic sea leading through the unstable fixed point to escape from the trap. Figure \ref{fig:0_0008-60-Rf} shows the phase space with only a small central regular island remaining, for a larger $q_5$. In this figure also the micromotion dressing of the pseudopotential motion is visualized.

To gain more understanding of the motion within the rf potential, we can use a correspondence between the phase space of the time-dependent rf potential and a higher dimensional phase space of a time-independent potential (see e.g. \cite{struckmeier2002canonical} and references therein). It is a general method that relates also to approaches for treating dynamical modes of periodic Hamiltonians \cite{rfions,rfmodes,lizuain2017dynamical}.
The motion generated by the Hamiltonian 
\be H\pare{z,p_z,t}=p_z^2/2+V_{\rm{rf}}^{\rm{1D}}\pare{z,t},\label{Eq:H1Drf}\ee
 can be embedded into a larger, 4D time-independent phase space, using the correspondence
\begin{equation}
H\pare{z,p_z,t} \rightarrow \mathcal{H}\pare{q,p,Q,P}=H\pare{q,p,Q}+P.\label{eq:HtoH}
\end{equation}
In \app{App:Embedding4D} we show that this embedding reproduces the original motion of $\{z(t),p_z(t)\}$ for a fixed energy shell of $\mathcal{H}\pare{q,p,Q,P}=0$. Since the conserved energy of a 4D time-independent Hamiltonian defines a 3D manifold, this justifies our statement that the original motion takes place in an effectively 3D phase space. 

Using this embedding, we can deduce that the phase space of the rf potential has a further important property. Since this phase-space is effectively 3D, any 2D invariant torus within it divides the space into two disconnected regions -- `inside' and `outside'. This means that it is possible to define the `last unbroken torus', that is the torus with maximal volume. In the stroboscopic map, it is simply the `last' continuous invariant curve, having maximal area. Since phase-space trajectories cannot cross, any motion started with an initial condition within the last unbroken torus will remain bounded. Even chaotic motion, within this torus, is disconnected from the chaotic `sea' outside. The existence of a torus that divides the phase space into two disconnected components facilitates a rigorous characterization of the trap.

\subsection{Trap characterization} \label{sec:psqevolution}

As a quantitative measure of the trap's ability to trap particles we
take the phase space area within the last unbroken torus. This measure
is in fact proportional to the action of the canonical action-angle
coordinates for integrable motion. It is a phase space invariant conserved in time, which is well defined both for the pseudopotential and the rf potential (for which the energy, in contrast, is not conserved), and allows an accurate comparison of the capability of the trap to store ions. The action is also monotonous with the ion's maximal kinetic energy (\fig{fig:StrobPlots}). Beyond the last unbroken torus the chaotic region is connected to the unstable fixed point and the ion is likely to quickly escape the trap. 

\begin{figure}
\includegraphics[width=0.44\textwidth]{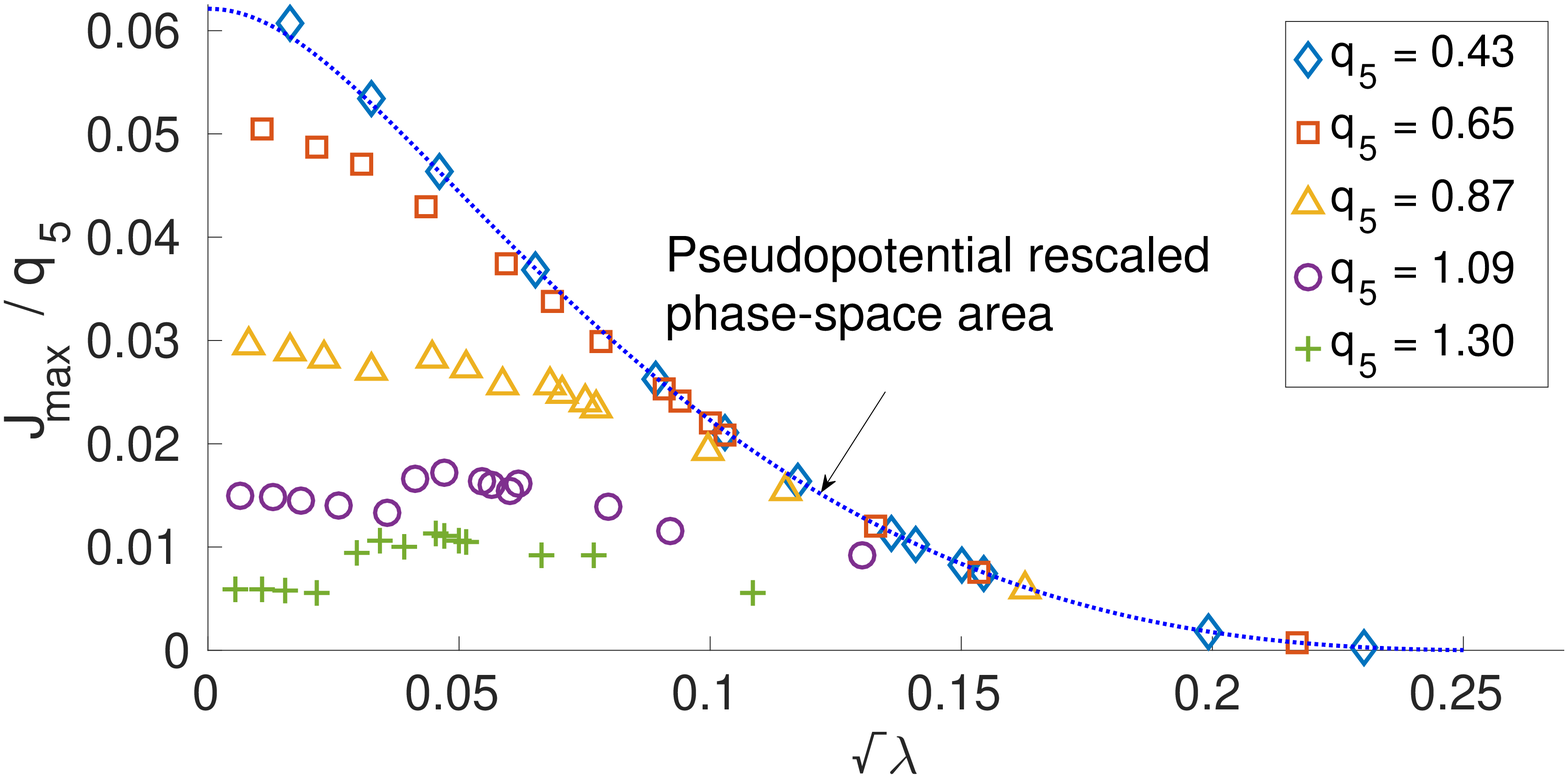}
\caption{Rescaled (non-dimensional) trapping phase space area, plotted as function of $\sqrt\lambda\propto \omega_x/U_{\rm rf}$, at increasing values of $q_5$ (see text for details, also \eq{eq:coordandtimerescale}  and \eq{eq:physicalparams}). The points are obtained by numerical simulation of the equations of motion in the rf potential, and the dotted curve is obtained for the pseudopotential from \eq{eq:Jqquadra}. Up to $q_5\approx 0.5$ ($U_{\rm rf}\approx 25$V), the pseudopotential gives a very good approximation of the global trap characteristics in the rf potential, with all simulated trap parameters falling very close to the pseudopotential curve, a function only of $\lambda$. The sharp jumps for larger $q_5$ values are due to the last unbroken torus breaking and being replaced by a nested one.}
\label{fig:JqzvssigmaPlots}
\end{figure}

For the pseudopotential, the phase space area within a given closed invariant curve can be easily calculated by integrating along the curve,
\begin{equation}
J=\oint p_zdz=2\int \sqrt{2\left[E-{V}_{\rm{pseudo}}^{1{\rm D}}(z)\right]}{d}z,
\label{eq:Jqquadra}
\end{equation} 
where the integration on the right hand side is to be performed from the minimal $z$ to the maximal $z$ of the closed curve, and $E$ is the energy of the particle. The phase space area of the rescaled pseudopotential of \eq{1paramPseudoPot1D} is given by $J_{\lambda}=J/q_5$, coming from \eq{tq_5rescale}.
The trapping phase space area, $J_{\rm{max}}$, is defined for each set of parameters as the maximal value of $J$ for the bounded motion. In fact it is the area bounded by the separatrix -- the closed curve that passes through the unstable fixed point.

\begin{figure}
\includegraphics[width=0.44\textwidth]{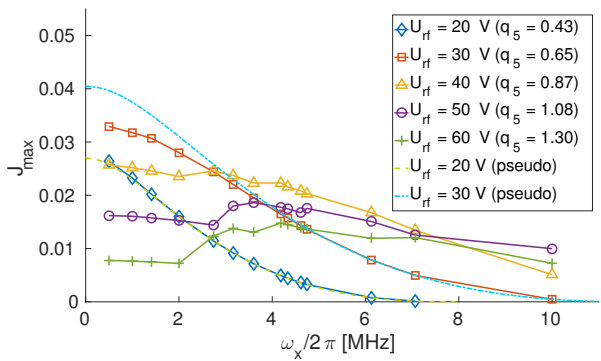}
\caption{Trapping phase space area $J_{\rm max}$ (non-dimensional), as a function of the axial trapping frequency $\omega_x$ for increasing values of $U_{\rm{rf}}$ (equivalently, $q_5$, see text for details). For $U_{\rm{rf}}=20\,$V, the results of the rf potential simulations overlap the pseudopotential curve. With $U_{\rm{rf}}=30\,$V, $J_{\rm max}$ increases for both the pseudopotential and the rf potential, but is lower within the rf potential for $\omega_x/2\pi \lesssim 3\,$MHz as compared with the pseudopotential curve. For higher values of $U_{\rm{rf}}$, the deviation becomes even more pronounced, since the pseudopotential trapping phase space area (not shown) only increases, while within the rf potential $J_{\rm max}$ decreases due to chaos. While for the pseudopotential $J_{\rm max}$ depends monotonously on $\omega_x$ and $U_{\rm{rf}}$, within the rf potential this dependence is non-monotonous and is generally more complicated due to the role played by chaotic motion. The maximal trapping is obtained with $U_{\rm rf}\approx 25\,$V and $\lambda\to 0$.}
\label{fig:JzvsomegaxPlots}
\end{figure}

Within the rf potential, the area within the last unbroken torus can be calculated directly from the stroboscopic map. However, this calculation is cumbersome when $p_z$ (restricted to positive values, say), is not a function of $z$, as can happen near resonance islands, see \fig{fig:StrobPlots}(d). We can use the fact that the stroboscopic map of $H$ coincides with a Poincar\'e surface of section of $\mathcal{H}$ defined in \eq{eq:HtoH}, as discussed in \app{App:Embedding4D}, and calculate the phase space area within a closed curve of the stroboscopic map using a method developed for calculating the action of 2D tori surfaces in the 4D phase space, described in \app{App:Action4D}.
With this method we calculate $J_{\rm max}$, the phase space trapping area, which is the area bounded by the last unbroken torus of the rf potential phase space.

To map the phase space of both the rf potential and the pseudopotential, we simulate the dynamics with initial conditions of $p_z\left(0\right)=0$ and small increments in $z(0)$ between $z_s$ and $z_u$. This allows us to cover the entire phase space (except some resonance islands), and also to construct numerically the transformation to action-angle variables throughout the region of regular motion. The simulation is repeated for different trap parameters, giving a complete characterization of the dependence on the parameters. An invariant curve is numerically identified  if the curve obtained from a single trajectory is continuous and has no width (to some accuracy). For the verification of the numerics, the pseudopotential simulations are compared with direct quadrature, numerically minimizing \eq{eq:dEdz} and integrating \eq{eq:Jqquadra}. The phase space trapping area is also calculated directly for convex curves from the stroboscopic map, and the tori winding number (defined in \app{App:Action4D}) is calculated and compared with its expected limit at the center of the trap, where $\nu_z$ is known.

\begin{figure}
\includegraphics[width=0.46\textwidth]{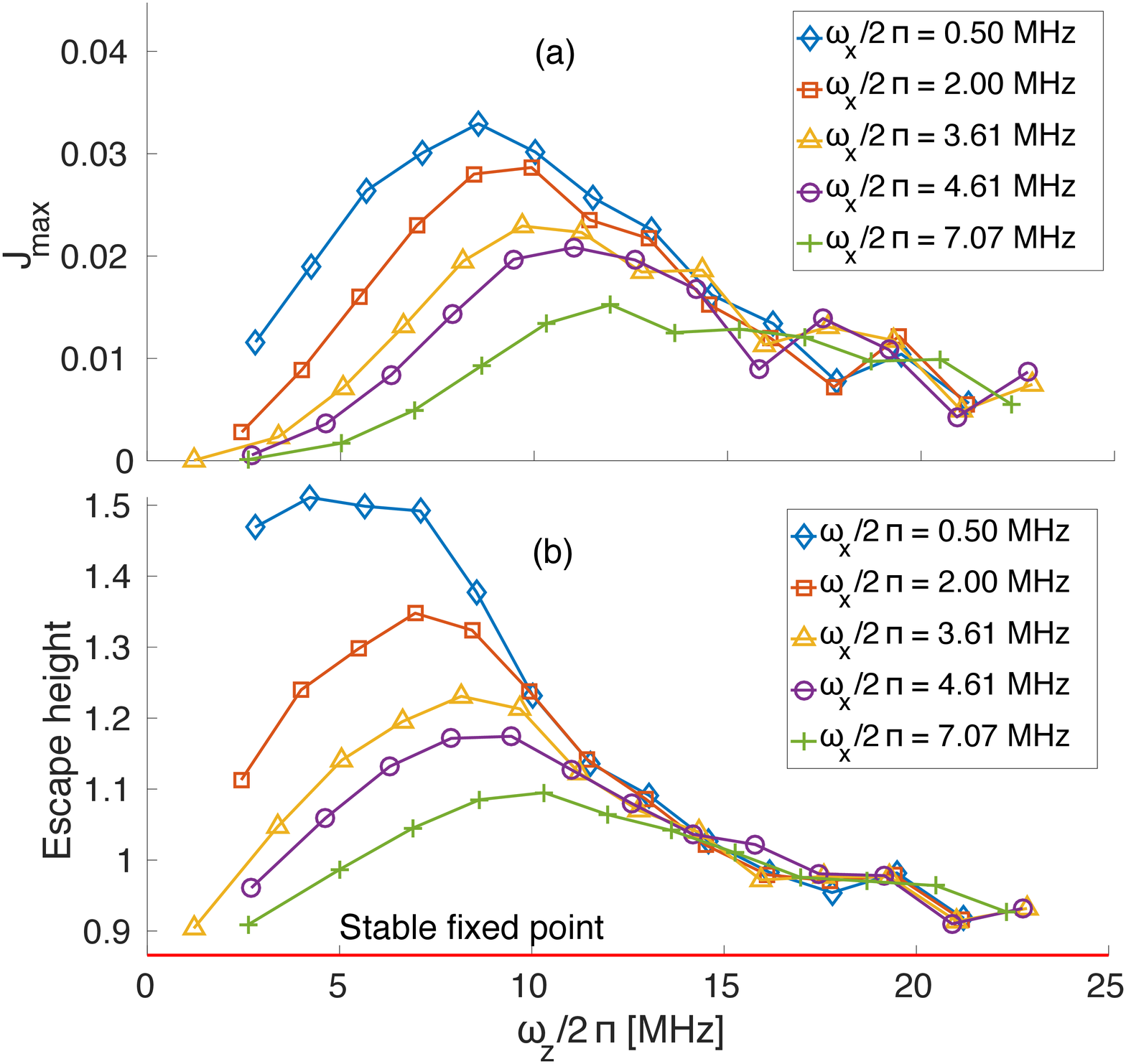}
\caption{(a) Trapping phase space area and (b) escape height [both in non-dimensional units, see \eq{eq:coordandtimerescale}], as a function of the linearized frequency $\omega_z$ at the stable fixed point, for increasing values of the axial trapping frequency $\omega_x$.
The global trapping characteristics are non-monotonic with $\omega_z$ and tend to become independent of $\omega_x$. In addition, it can be seen that the escape height and phase space trapping area do not reach their maximum together as a function of the parameters.}
\label{fig:JzvsomegazPlots}
\end{figure}
 
With the rescaled units of \eq{tq_5rescale},
\fig{fig:JqzvssigmaPlots} presents the complete information on the 3D
phase space of the symmetric, unbiased 5-wire trap. We find that the
pseudopotential measures the phase space trapping area of the rf
potential very accurately up to $q_5\approx 0.5$
($U_{\rm rf}\approx 25\,$V), {\it for any} $\lambda$. Further
  increasing $U_{\rm rf}$ (i.e.\ $q_5$) leads to two competing
  effects.  On the one hand, as for smaller values of $U_{\rm rf}$,
  the size of the trapped region within the pseudopotential
  approximation keeps increasing.  This again is due both to the fact
  that increasing $q_5$ decreases $\lambda$ which in turn increases
  the depth of the potential, and with it the maximal rescaled phase
  space trapping area $J_{\rm max}/q_5$, and also due to the fact that in
  the original unscaled phase space units, $J_{\rm max}$ grows
  linearly with $q_5$ (at fixed $\lambda$, i.e.\ at fixed scaled
  pseudopotential dynamics). However, at the same time, within the rf potential, the rf perturbation
  starts to significantly affect the nature of the dynamics: invariant
tori begin to break, and the effective trapping is lowered due the
increasing size of the chaotic regions connected to the exterior of
the trap. 
 The consequences of this mechanism can be observed for
  instance in \fig{fig:JqzvssigmaPlots} for  $q_5 \gtrsim 0.65$. Furthermore, at any fixed value of $q_5\gtrsim 0.87$
  ($U_{\rm{rf}}\gtrsim 40$V), $J_{\rm max}$
  can be seen in the same figure to make large jumps as a function of
  $\sqrt{\lambda}$ or of $\omega_x$. These large jumps can be
  associated with the breaking of a torus, leading to the connection
  of a chaotic layer around a resonance (island chain) with the main
  chaotic region coming from the unstable fixed point. The same
  phenomenon is visible in \fig{fig:JzvsomegaxPlots}, that is discussed below.

The competition between the increase of the pseudopotential phase
  space trapping area and the increase of the chaotic region leads to a
non-monotonic dependence of the phase space trapping area (in unscaled
units) on the parameters in the rf potential. This can be seen in
\figs{fig:JzvsomegaxPlots}-\ref{fig:JzvsomegazPlots}(a), where the
secular frequencies $\omega_x$ and $\omega_z$ serve as the control
parameters. To maximize the total phase space trapping area, taking
$U_{\rm rf}\approx 25\,$V, at the border of the validity of the 
pseudopotential, is optimal, and $\lambda\to 0$ (equivalently,
$\omega_x\to 0$) is required. Otherwise, if $\omega_x$ is constrained,
the maximal trapping will be obtained at different values of
$U_{\rm rf}$.
In experiments with trapped ions it is often desirable to have the
linearized trapping frequencies $\omega_{\alpha}$ as high as
possible. However, increasing $\omega_x$ causes anti-trapping in the
radial coordinates and increases $\lambda$ (at fixed
$\omega_z$). Increasing $\omega_z$ (at fixed $\omega_x$) requires
increasing $U_{\rm rf}$ and hence the non-integrable perturbation. Both
effects eventually degrade the trap.

Finally, in
\fig{fig:JzvsomegazPlots}(b) we plot the escape height from the
trap, defined as the minimal position $z>z_s$, at $p_z=0$ and
$t=0\pmod{\pi}$, for which the ion escapes the trap. The escape height and the phase space trapping area do
not attain their maximum together, which emphasizes the importance of
the trap's capability to trap high momentum particles, measured by the
phase space area.

\begin{widetext}

\section{Hamiltonian motion in two spatial dimensions}\label{sec:Hamiltonian2D}

The pseudopotential for the coupled $yz$ motion, including the bias voltage, can be decomposed as
\be V_{\rm pseudo}^{\rm 2D}=\frac{1}{2}a_z\left[y^2+\pare{z-z_{\rm{0}}}^2\right] + q_5^2 \left[V_{\rm avg}^{\rm 2D}\right] +a_5\left[V_{\rm 5w}(y,z)-\frac{1}{3}\right],\label{eq:Vpseudo2D}\ee
with $V_{\rm 5w}$ defined in \eq{eq:5wnew}, the factor of $1/3$ makes the potential vanish at its minimum, and $V_{\rm avg}^{\rm 2D}$, the time-independent approximation to the time-dependent part of \eq{V5wnondim}, can be derived as outlined in \app{app:pseudopot} and reads
\be  V_{\rm avg}^{\rm 2D}=\frac{16 \left(16 \left(y^2+z^2\right)^2+24 (y-z)
   (y+z)+9\right)}{\pi ^2 \left((1-2 y)^2+4 z^2\right)
   \left((3-2 y)^2+4 z^2\right) \left((2 y+1)^2+4
   z^2\right) \left((2 y+3)^2+4 z^2\right)}. \label{eq:Vavg2D}\ee
With the second rescaling of time ($t\to q_5 t$) given in \eq{tq_5rescale} we get the rescaled pseudopotential
\be V_{\lambda}^{\rm 2D}=-\frac{1}{2}\lambda\left[y^2+\pare{z-z_{\rm{0}}}^2\right] + V_{\rm avg}^{\rm 2D} - \lambda_{\rm b}\left[V_{\rm 5w}(y,z)-\frac{1}{3}\right],\label{eq:Vlambda2D}\ee
\end{widetext}
where we have defined the pseudopotential bias parameter,
\be \lambda_{\rm b}=-\frac{a_5}{q_5^2}\propto -\frac{U_{\rm b}}{(U_{\rm rf})^2}.\label{Eq:lambda_bias}\ee



Let us first consider the unbiased potential ($\lambda_{\rm b}=0$). By running careful numerical simulations of the dynamics over a fine grid of initial conditions, we have found almost no signatures of chaos throughout the phase space up to the unstable fixed point.
Pinning down the reason for this near-integrability and identifying the implied additional constant of motion is beyond the scope of the current work. 
However, this integrability implies that any added perturbation, if not too large, introduces only a limited structural change of the phase space.
In this section we demonstrate this statement for two separate perturbations of the integrable pseudopotential, and then their combined effect. First we consider (in \seq{Sec:Pseudo2D}) the effect of a negative bias voltage on the structure of the pseudopotential phase space. Then  in \seq{Sec:rf2D} we study the full rf potential, which can be considered as a second perturbation scaled by $q_5$ that is added to the pseudopotential, including the bias.

\subsection{The 4D pseudopotential phase space}\label{Sec:Pseudo2D}

 In the 4D phase space of the $yz$ pseudopotential, increasing $\lambda_{\rm b}$ increases the trap depth by pushing the unstable fixed point $z_u$ up, and eventually completely eliminating it, allowing the ion to be trapped without escaping for any energy. At the same time, this perturbation gradually deforms the phase space, breaks the main island into several smaller islands, and introduces a chaotic region where the ion is expected to be more susceptible to heating and can escape from the trap (within the rf potential, to be studied in \seq{Sec:rf2D}). Hence the bias voltage induces a competition between two mechanisms of opposite nature, that we investigate in the following.

We have simulated the equations of motion derived from \eq{eq:Vpseudo2D}, varying the parameters. We vary $\lambda$ between $\surd{\lambda}\approx 0.016$ and $\surd{\lambda}\approx 0.065$ [$\omega_x=2\pi\times 0.5\,$MHz to $\omega_x=2\pi\times 2\,$MHz at $U_{\rm rf}=20\,$V], corresponding to the first quarter of the horizontal axis of \fig{fig:JqzvssigmaPlots}. We vary $\lambda_{\rm b}$ between $\surd{\lambda_{\rm b}}=0.21$ [$U_{\rm{b}}=-0.2\,$V at $U_{\rm rf}=20\,$V] and $\surd{\lambda_{\rm b}}=0.43$ [$U_{\rm{b}}=-0.8\,$V at $U_{\rm rf}=20\,$V], in addition to the unbiased potential $\lambda_{\rm b}=0$.

\begin{figure}[!t]
\includegraphics[width=3.4in]{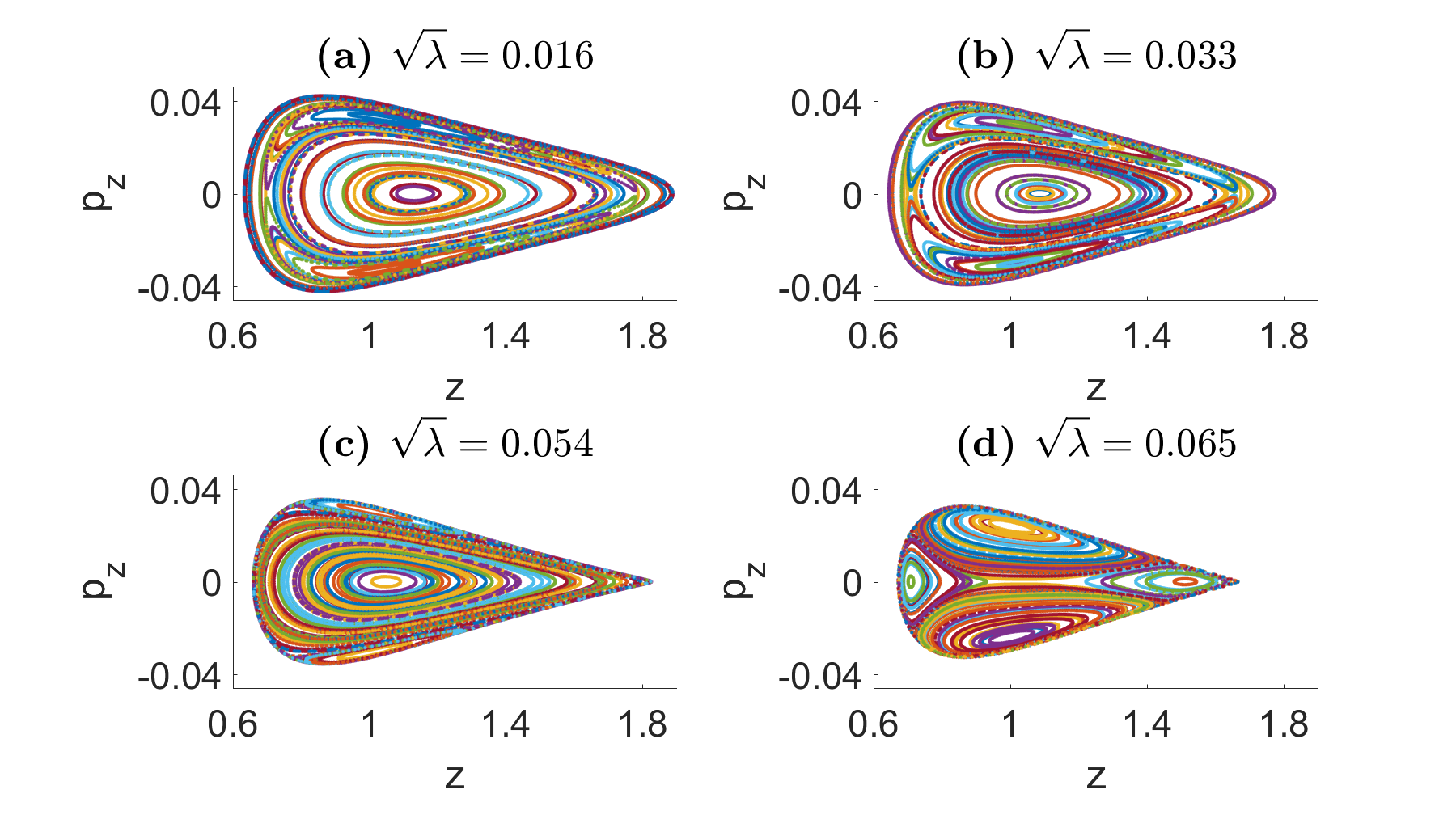}
\caption{Poincar\'e surfaces of section simulated using the biased pseudopotential of \eq{eq:Vpseudo2D} with a low bias voltage, $\surd{\lambda_{\rm b}}=0.21$ [$U_{\rm{rf}}=20\,$V, $U_{\rm{b}}=-0.2\,$V]. In each panel (a)-(d), a different value of $\lambda$ is indicated, corresponding to $\omega_x=2\pi\times\left\{0.5,1,1.66,2\right\} {\rm MHz}$ respectively, and the energy shell is close to the maximal simulated energy value for the respective $\lambda$ (and $\lambda_{\rm b}$). For the parameters of panels (a)-(b) the ion does not escape for any initial condition in the simulated energy shells [constrained by \eq{eq:z0max}], while for panels (c)-(d) [with increased $\lambda$], the unstable fixed point $z_u$ is nearby in energy. We find that for this relatively small value of  $\lambda_{\rm b}$, $z_u$ can be pushed significantly up, while the phase space remains mostly regular (albeit with large resonance islands). We note that in contrast to \figs{fig:StrobPlots}-\ref{fig:0_0008-60-Rf}, here the island centers do not correspond to any fixed point. In addition, each island is seen to be  a section of a different, disconnected family of tori.}
\label{fig:PS2D0}
\end{figure}

 For each pair $(\lambda,\lambda_{\rm b})$ we run the simulation on a grid of energies between 0 and up to a maximal energy, which we take to be the energy of a particle at rest at a certain height $z_{\rm max}^0$. For $\lambda_{\rm b}=0$, this height is just $z_{\rm max}^0= z_u$. For $\lambda_{\rm b}>0$, $z_u$ is pushed up significantly, and we truncate the simulations at an arbitrary energy shell corresponding to a particle at rest at a height of
\be z_{\rm max}^0\approx 2.3\times z_{\rm s}\approx 2.\label{eq:z0max}\ee
As we discuss in detail in the conclusions of the following
subsection, trajectories that reach very high above the electrode are
expected to be less relevant from an experimental point of view, and
hence the condition above is a reasonable one.

We begin by examining the structure of the phase space as a function of $\lambda$ and $\lambda_{\rm b}$. For a small value of the bias potential we find that hardly any parts of phase space become chaotic as $z_u$ is pushed up. Figure \ref{fig:PS2D0} demonstrates this for different values of $\lambda$. Each panel shows a Poincar\'e surface of section, constructed by taking simulations with different initial conditions at a fixed energy, and plotting a point in the plane of $\{z,p_z\}$, every time a simulated trajectory goes through $y=0$, with $p_y>0$. 
Except for the largest $\lambda$ value, the phase space contains a large central regular island. For the two highest $\lambda$ values shown, the cusp in the trajectories indicates the proximity of the unstable fixed point in a higher energy shell. For the maximal value of $\lambda$, the phase space is completely broken into smaller resonance islands, although the motion still stays almost regular. 

\begin{figure}[!t]
 \includegraphics[width=3.4in]{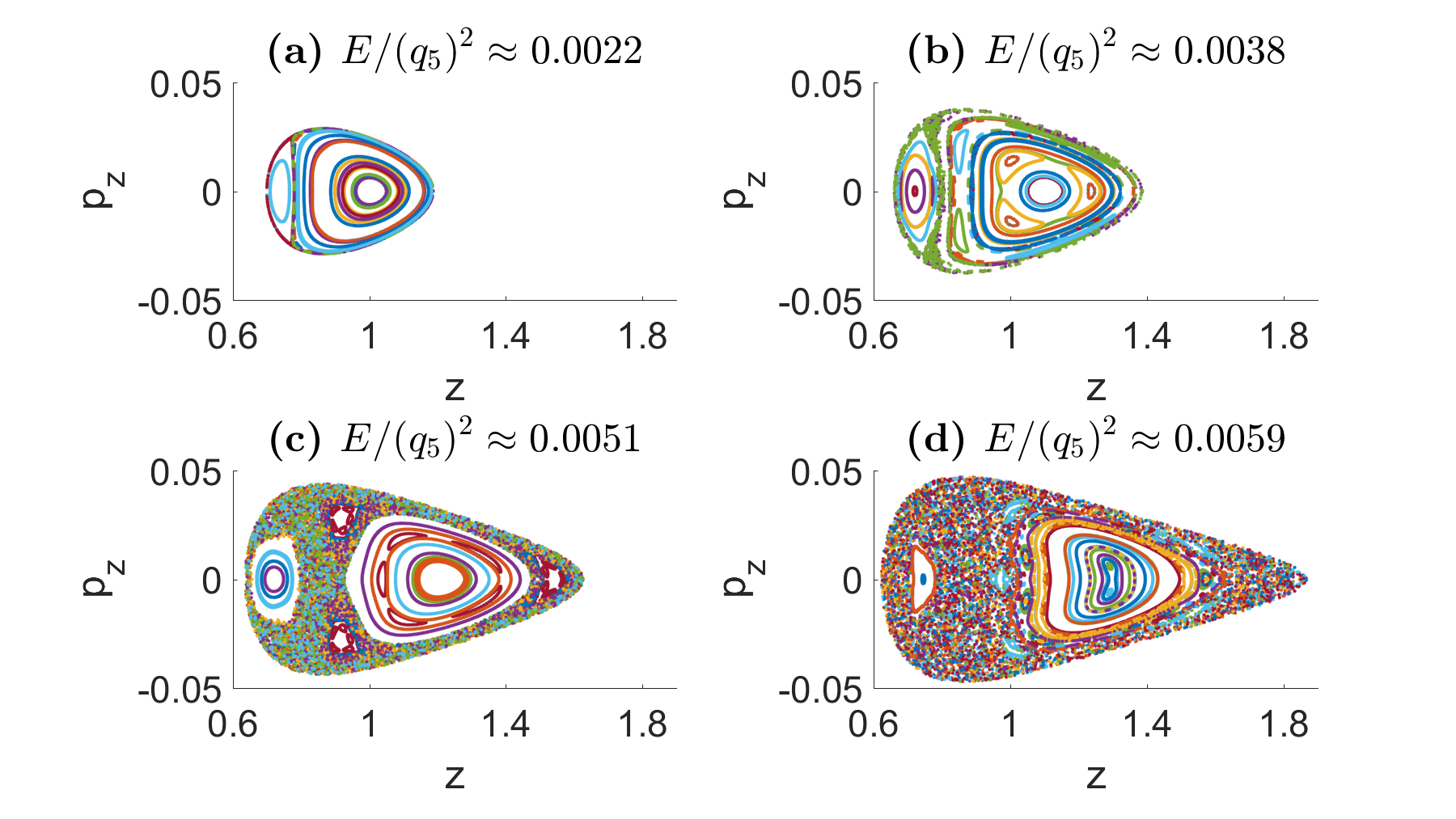} \caption{Poincar\'e surfaces of section at increasing values of the energy within the biased potential of \eq{eq:Vpseudo2D} with $\surd\lambda=0.065$ and $\surd{\lambda_{\rm b}}=0.30$ [$U_{\rm{rf}}=20\,$V, $U_{\rm{b}}=-0.4\,$V, and $\omega_x=2\pi\times 2\,{\rm MHz}$]. The phase space expands, breaks at first into two main islands, and becomes more chaotic at higher energies where one island is almost destroyed. The entire chaotic region and even parts of the islands will gradually lead to  escape out of the trap within the rf potential as its amplitude is increased beyond a threshold (see \figs{fig:Volume2D1}-\ref{fig:Volume2D2}).} \label{fig:PS2D1} \end{figure}

\begin{figure}[t]
\includegraphics[width=3.4in]{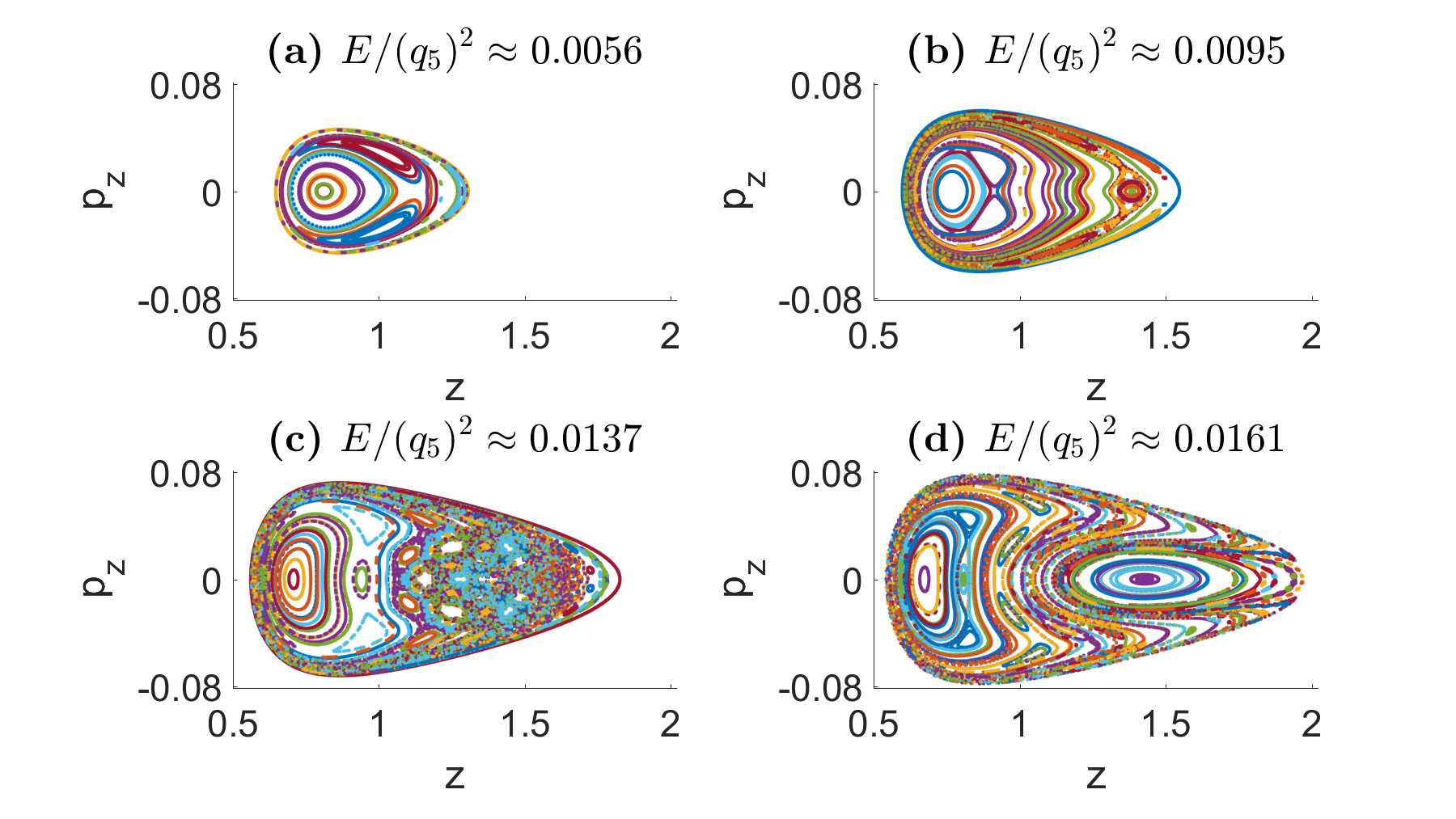}
\caption{Poincar\'e surfaces of section as in \fig{fig:PS2D1}, with $\surd\lambda=0.016$ and $\surd{\lambda_{\rm b}}=0.43$ [$U_{\rm{rf}}=20\,$V, $U_{\rm{b}}=-0.8\,$V, and $\omega_x=0.5\pi\times 2\,{\rm MHz}$]. Here the central island is pushed to the lower $z$ side, while at higher $z$ values the motion becomes at some high energy almost completely chaotic and then a new island emerges with a large size. We note the different scale of the axes (in particular of $p_z$) as compared with \figs{fig:PS2D0}-\ref{fig:PS2D1}. The relatively small value of $\lambda$ allows increasing $\lambda_{\rm b}$ and significantly enlarging the phase space while still maintaining large (nearly) regular islands at high energies.}
\label{fig:PS2D2}
\end{figure}

Fixing the parameters, with $\lambda$ at the maximal value we have used and $\lambda_{\rm b}$ at an intermediate value, \fig{fig:PS2D1} shows Poincar\'e sections at different, increasing energies. The phase space splits into two central islands with an increasing chaotic sea between them. In \fig{fig:PS2D2}, for the largest value of $\lambda_{\rm b}$, a similar phase space splitting occurs (with the main island centered at lower $z$ values), together with island destruction and an emergence of a new island. However, since in \fig{fig:PS2D2} the value of $\lambda$ is small, even in the high energy shells there can be a large fraction of (nearly) regular motion -- and we note that the scale of the $p_z$ axis is $\sim 1.6$ times that of \fig{fig:PS2D1} and twice that of \fig{fig:PS2D0}.

In order to quantify the trap's effectiveness within the
pseudopotential approximation, we single out a 4D phase space volume
with a given property, to be defined below. Since such a volume is a
function of the conserved energy $E$ up to which it is measured, it
can be obtained by integration of a 3D volume of interest
$\Lambda^{\rm (3D)}_\chi(E)$ over the energy shells, 
\begin{align}
 \Lambda^{\rm (4D)}_\chi(E) & = \int_0^E  dE' \left[\Lambda^{\rm
    (3D)}_\chi(E')\right] , \\
\Lambda^{\rm   (3D)}_\chi(E) & = \int d\vec q d\vec p \chi(\vec q,\vec p) \delta(E - H(\vec q,\vec p)),
\end{align}
with $\vec q = \{y,z\}$, $\vec p = \{p_y,p_z\}$, $H$ the pseudopotential Hamiltonian, and $\chi(\vec q,\vec p)$ an indicator function (taking therefore either  the value 0 or 1) that selects some region of phase space with a specific property (different $\chi$ functions will be introduced below).
  We note that by its definition, $ \Lambda^{\rm (3D)}_\chi(E)$ is independent of the rescaling of time [in \eq{tq_5rescale}], leading to the pseudopotential of  \eq{eq:Vlambda2D}, but $ \Lambda^{\rm (4D)}_\chi(E)$ is not. 

  Within each energy shell, $\Lambda^{\rm (3D)}_\chi(E)$ can be
  calculated in a straightforward manner from the Poincar\'e section, as
  outlined in \app{App:Volume4D}. 
We stress here that by its definition, the Poincar\'e section contains
the full information on the motion (at each energy). Within the
pseudopotential approximation, the Poincar\'e section is 2D, which
simplifies the analysis significantly, since we can reduce the 4D
indicator functions $\chi(\vec q,\vec p)$ to their 2D counterparts
$\chi(z,p_z)$ within the  Poincar\'e section.

 We introduce two indicators, $\chi_{\rm t}(z,p_z)$ and
  $\chi_{\rm r}(z,p_z)$, and assign their value within each Poincar\'e
  section (gridded appropriately).  For the first, we set
  $\chi_{\rm t}(z,p_z)=1$ on every point of the Poincar\'e section
  such that the corresponding motion is bounded within the
  pseudopotential approximation. Hence $\chi_{\rm t}$ measures the
  total pseudopotential phase space volume. We further define
  $\chi_{\rm r}(z,p_z)$ to equal 1 on points of the
  Poincar\'e section that (in the pseudopotential approximation) are
  {\em within} the last unbroken torus of the center islands (around
  $p_z=0$), and 0 otherwise. With this definition, it is relatively
  fast to compute $\chi_{\rm r}$ numerically, and it gives a good
  approximation of the regular motion volume. A more accurate
quantification where regular regions outside the last unbroken torus
are included and chaotic regions within the last unbroken torus are
excluded would be significantly more complicated to implement for very extensive
simulations, and not necessarily more relevant.

Figure \ref{fig:Volume2D0} compares these two measures of the
pseudopotential phase space volume, as a function of $\lambda$,
$\lambda_{\rm b}$. The limit of integrability appears to be near
$\lambda\to 0$ and $\lambda_{\rm b} \to 0$. For the smallest $\lambda$
value, even a large $\lambda_{\rm b}$ (up to the largest one) allows a
large gain of (mostly) regular phase space
(\fig{fig:PS2D2}). Likewise, for the smallest value of $\lambda_{\rm
  b}$, the phase space can change its structure but maintain primarily
regular motion up to a large value of $\lambda$
(\fig{fig:PS2D0}). However, as both ${\lambda}$ and ${\lambda_{\rm
    b}}$ increase beyond their minimal values, the increased
{useful} phase space volume due to the bias may be saturated at a certain ($\lambda$-dependent) $\lambda_{\rm b}$ value, by the destruction of the regular motion regions that come with it (\fig{fig:PS2D1}). Within the pseudopotential, the dynamics are independent of the value of $U_{\rm rf}$, which only rescales the units as discussed below. In \fig{fig:Volume2D0}, $U_{\rm rf}=20\,$V is chosen for comparison with the dynamics within the rf potential, studied in the following subsection.
  
\subsection{The 5D rf phase space}\label{Sec:rf2D}

\begin{figure}[!t]
\includegraphics[width=3.4in]{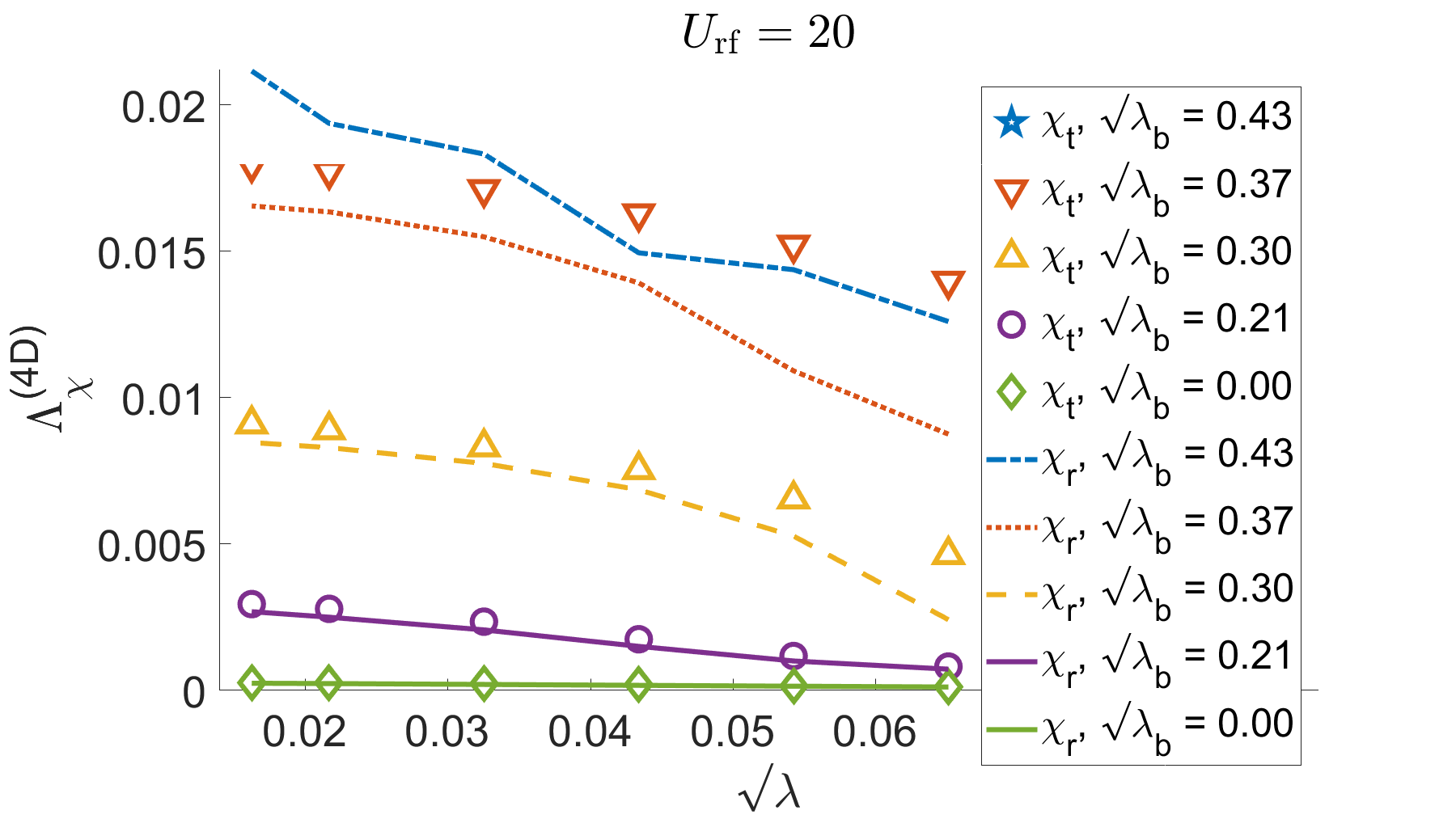}
\caption{The 4D phase space volume $\Lambda^{\rm (4D)}_\chi$,
  measuring pseudopotential trajectories with a given property, as
  defined by two different indicator functions. The symbols correspond
  to the total pseudopotential volume ($\chi_{\rm t}$), and the lines
  to motion bounded by the last unbroken torus of centred islands, approximately regular ($\chi_{\rm r}$). The values are given for $U_{\rm rf}=20\,$V for comparison with \figs{fig:Volume2D1}-\ref{fig:Volume2D2}, but within the pseudopotential the dynamics is fully determined by $\lambda$ and $\lambda_{\rm b}$ and the volume $\Lambda^{(4D)}_{\chi}$ scales with $(U_{\rm rf})^2$. The open pentagram symbols corresponding to $\surd \lambda_{\rm b}=0.43$ are outside of the plot scale (at $\Lambda^{(4D)}_{\chi}\sim 0.03$), which is maintained for comparison with \figs{fig:Volume2D1}-\ref{fig:Volume2D2}. See the text for a detailed discussion.}
\label{fig:Volume2D0}
\end{figure}

Turning to the full $yz$ rf potential of \eq{V5wnondim}, beyond the
addition of a third parameter ($U_{\rm rf}$), the motion takes place
in an effectively 5D phase space. The construction of Poincar\'e
sections in this case is not straightforward, and in addition a torus
of regular motion (which is only a 3D manifold), cannot divide the
phase space into disconnected regions, so there is no longer a {\it
  last} unbroken torus. 
 For any amplitude of the perturbation, there will always be some (possibly exponentially small) measures of tori destroyed by the perturbation. Since no regular torus divides the phase space, these broken tori become connected and 
  transport between different regions of phase space can then proceed in a complex manner.
     A comprehensive analysis becomes thus
significantly more complicated. 

{  In particular, within a 5D phase space (and above), Arnol'd diffusion is a possible transport mechanism \cite{lichtenberg1992regular}. This process takes place through successive hops from one thin chaotic  layer to another one.  In practice, due to various uncontrollable perturbations in ion traps (from e.g., fluctuations of voltages and charges on the electrodes, or stray electromagnetic fields), the ion is constantly heated and slowly diffuses. The strength of the random forces is independent of the phase space structure (regular or chaotic) and can be expected to overwhelm the exponentially suppressed Arnol'd diffusion rate (from deep within the regular regions).}
In this subsection we therefore take a
simpler approach, in which we use information about the dynamics from
simulations with the full rf potential, but we do not analyze the full
5D phase space of the rf motion, and rather keep the simplicity
of being able to use the 2D Poincar\'e sections constructed for the 4D
phase space of the pseudopotential.

\begin{figure}[!t]
\includegraphics[width=3.4in]{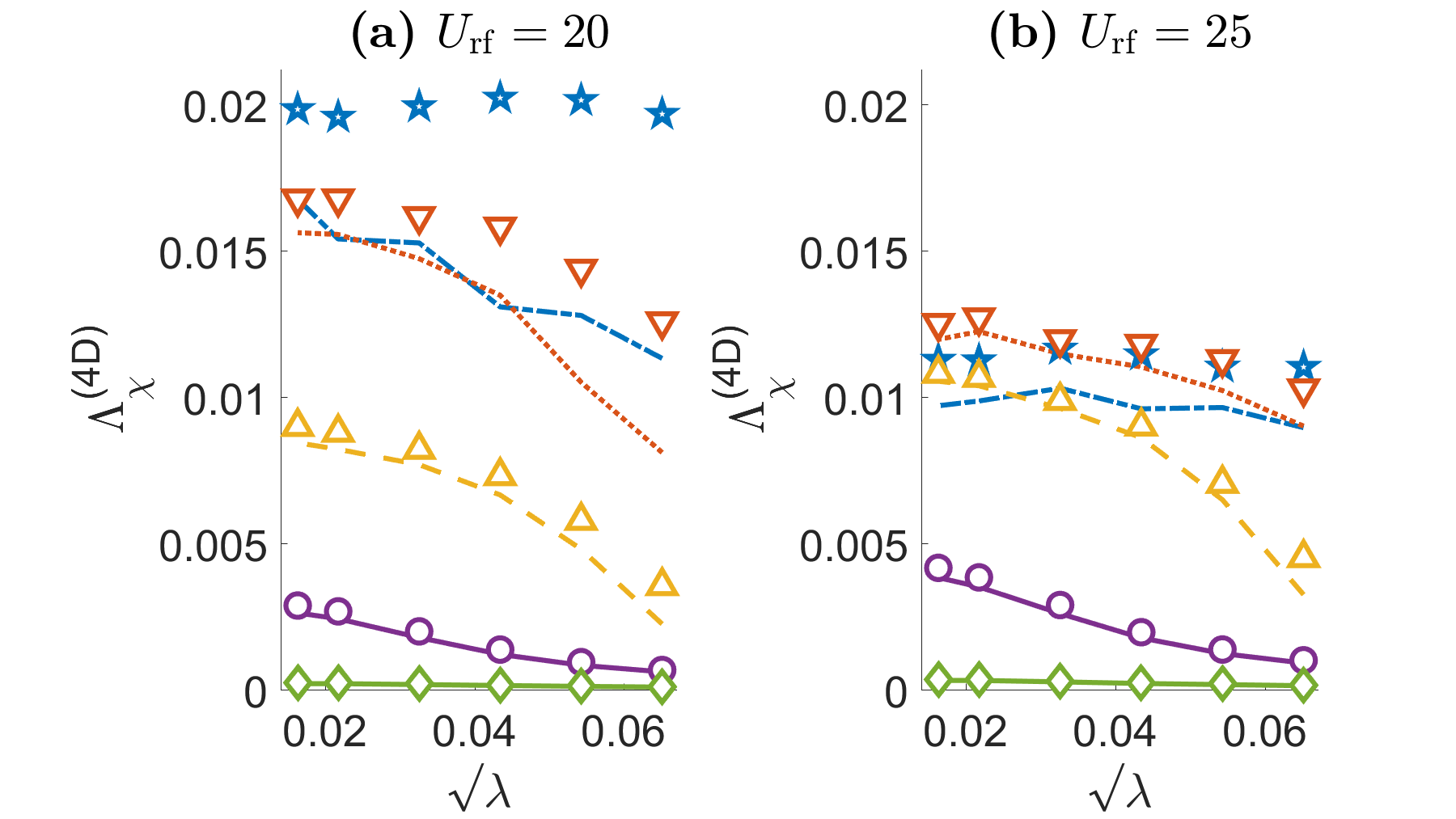}
\caption{The 4D phase space volume $\Lambda^{\rm (4D)}_\chi$, measuring pseudopotential trajectories with a given property, that are non-escaping with the rf potential. The symbols correspond to all pseudopotential trajectories, and the full lines to  those bounded by the last unbroken torus of center islands. See \fig{fig:Volume2D2} for the legend of $\lambda_{\rm b}$ values, and the text for a detailed discussion. The maximal (mostly) regular trapped volume which can be considered as robust to non-integrable perturbations is obtained with $U_{\rm rf}=20\,$V, $\lambda\to 0$, and $\surd{\lambda_{\rm b}}\approx 0.37$.}
\label{fig:Volume2D1}
\end{figure}

Repeating the same set of simulations as described in the
previous subsection (for the pseudopotential), within the full $yz$ rf
potential of \eq{V5wnondim}, we take $U_{\rm rf}$ between $20\,$V and
$35\,$V (corresponding to $q_5\in [0.43,0.76]$), as suggested by the
analysis of \seq{Sec:Hamiltonian1D}.  At each value of $U_{\rm rf}$,
we simulate exactly the same initial conditions as simulated
with the pseudopotential. We run the simulation for 20,000 rf periods,
a duration longer than the typical lifetime of the trajectories that were
observed to escape. In addition to $\chi_{\rm t}$ and $\chi_{\rm r}$
that characterize the motion solely within the pseudopotential, we can
now define $\chi_{\rm n}(z,p_z)=1$ on every point of the {\em
  pseudopotential} Poincar\'e section if it belongs to a trajectory
that does not escape from the trap within the {\it rf potential}
simulation, when initiated at time $t=0$ with the same initial
condition (and 0 otherwise).

\begin{figure}[!t]
\includegraphics[width=3.4in]{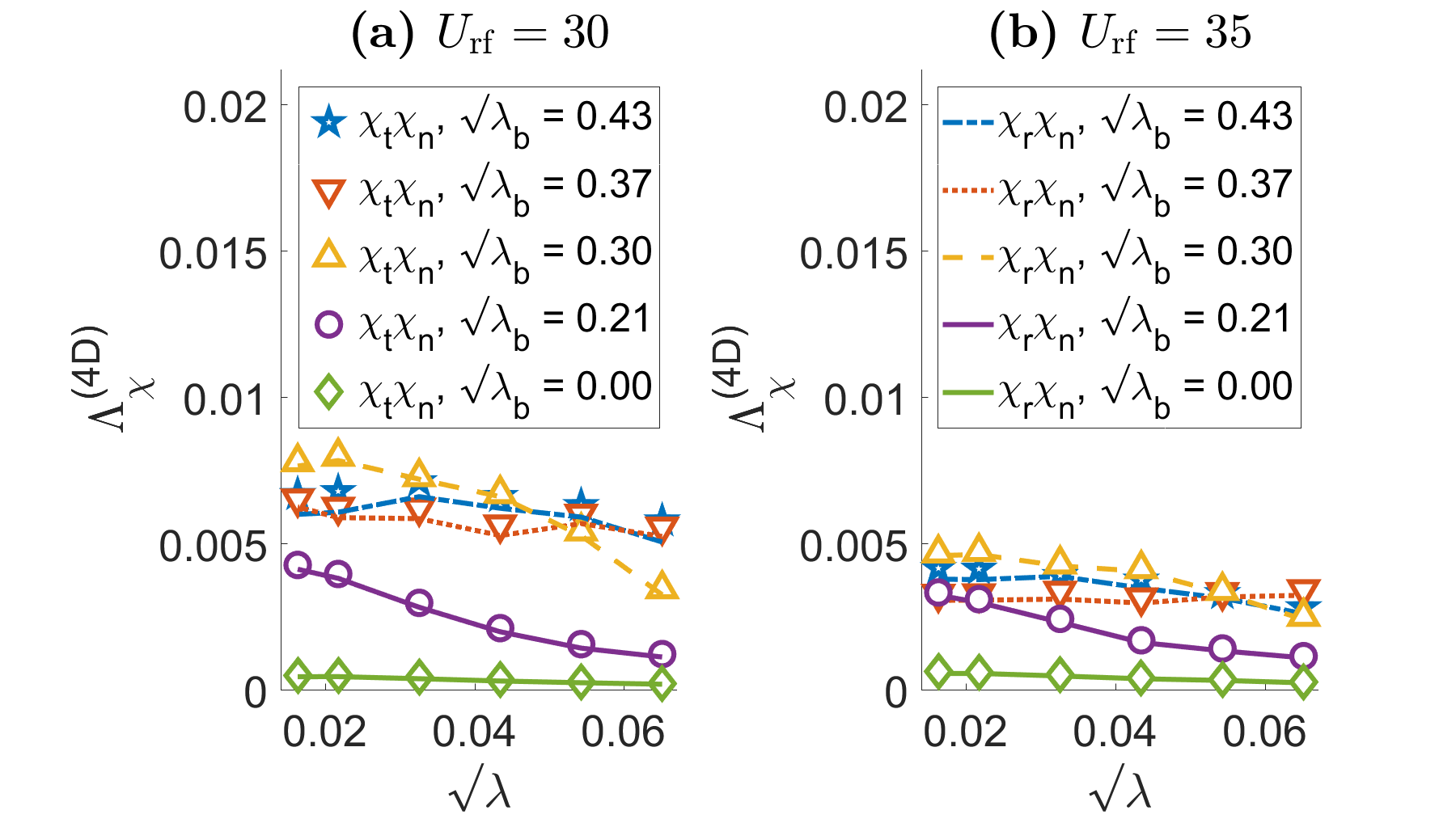}
\caption{As in \fig{fig:Volume2D1}, but with larger $U_{\rm rf}$ values. The legend in panel (a) shows the value of $\lambda_{\rm b}$ assigned to the symbols representing the total pseudopotential phase space volume not escaping within the rf potential ($\chi_{\rm t}\chi_{\rm n}=1$), and the legend in panel (b) gives $\lambda_{\rm b}$ assigned to the full lines showing the pseudopotential phase space volume (mostly) regular and bounded within the center islands, also non-escaping ($\chi_{\rm r}\chi_{\rm n}=1$).}
\label{fig:Volume2D2}
\end{figure}

In \figs{fig:Volume2D1}-\ref{fig:Volume2D2} we
compare, as a function of $\lambda$, $\lambda_{\rm b}$, and $U_{\rm
  rf}$, the total pseudopotential phase space volume non-escaping
within the rf potential ($\chi_{\rm t}\chi_{\rm n}$), with the
pseudopotential volume bounded within the last unbroken torus of
center islands, and non-escaping within the rf potential ($\chi_{\rm
  r}\chi_{\rm n}$). Although within the pseudopotential, the dynamics
are determined by $\lambda$ and are identical for any value of $U_{\rm
  rf}$ at fixed $\lambda$, the 4D phase space volume $\Lambda^{\rm
  (4D)}_\chi$ in fact scales $\propto (U_{\rm rf})^2$. This scaling
would imply an increase of the volume with $(U_{\rm rf})^2$ for fixed
$\lambda$ and $\lambda_{\rm b}$. However the competition with the
effect of trajectories gradually escaping from the trap due to the
increase of $U_{\rm rf}$, in particular in the higher energy shells
and larger values of $\lambda$ and $\lambda_{\rm b}$, determines the
effective gain due to the increase of $U_{\rm rf}$.

By comparing \fig{fig:Volume2D0} with \fig{fig:Volume2D1}(a) we find that for  intermediate parameter values, $\surd{\lambda}\lesssim 0.043$ and $\surd{\lambda_{\rm b}}
\lesssim 0.30$,
almost the entire pseudopotential phase space volume is bounded within unbroken tori, and hardly any of that escapes within the rf simulation at $U_{\rm rf}=20\,$V. 
For larger $\lambda$, ${\lambda_{\rm b}}$ values, there could be a large chaotic volume that remains trapped. 
Comparing with panel (b), we see that the trapped volume increases (and is mostly regular) if increasing $U_{\rm rf}$ to 25\,V for $\surd{\lambda_{\rm b}}\lesssim 0.3$, while for larger $\lambda_{\rm b}$, we see a cross-over from an increase to a decrease of the volume (depending on $\lambda$), due to trajectories escaping within the rf potential. Figure \ref{fig:Volume2D2}(a) shows that for $U_{\rm rf}=30\,$V and $\surd{\lambda_{\rm b}}\gtrsim 0.3$, trajectories continue to escape from the rf potential and lead to a decrease of trapped phase space volume as compared with $U_{\rm rf}=25\,$V, since even the regular parts shrink. For $U_{\rm rf}=35\,$V [\fig{fig:Volume2D2}(b)] even for $\lambda_{\rm b}\gtrsim 0.2$ the volume decreases (as compared with $U_{\rm rf}=30\,$V).

Hence, we see again how the near-integrable pseudopotential behaves under the addition of combined  perturbations. The pseudopotential volume bounded by the last unbroken torus of center islands appears to give a good measure of the trajectories that are robust to escape under the rf potential.
It indicates the trap's capability to trap particles of a high kinetic energy, and it is a relatively simple quantity to calculate numerically. 
The border of validity of the pseudopotential (for approximately regular motion) was found in \seq{Sec:Hamiltonian1D} to be $q_5\approx 0.5$ ($U_{\rm rf}\approx 25\,$V), for any value of $\lambda$. We can conclude that this threshold holds also in the coupled $yz$ motion, restricted to intermediate values of $\lambda$ and $\lambda_{\rm b}$. Above this $U_{\rm rf}$ threshold, all chaotic trajectories escape, and the regular region shrinks as well. For lower values of $U_{\rm rf}$, even parts of the chaotic motion could remain trapped (for very long times), but for $\surd{\lambda}\lesssim 0.043$ and $\surd{\lambda_{\rm b}}\lesssim 0.30$, the phase space is primarily regular. For higher values of $\lambda$ or $\lambda_{\rm b}$ the picture is more complicated and sensitive to an interplay of the parameters, as described above.

Figure \ref{fig:Volume2D1} also allows us to deduce how to maximize a trapped phase space volume that is (mostly) regular. A value of  $U_{\rm rf}$ that is just below the threshold of pseudopotential validity, is apparently close to optimal. At lower $U_{\rm rf}$ the phase space volume shrinks because of the energy scaling, and at higher $U_{\rm rf}$ significantly more trajectories escape. For the trap parameters of \eq{eq:physicalparams}, it is $U_{\rm rf}\approx 20\,$V. Taking $\lambda\to 0$ is required, and $\surd{\lambda_{\rm b}}\approx 0.37$ can be considered to give the maximal volume which is robust to small integrability-breaking perturbations. Although increasing $\lambda_{\rm b}$ allows according to panel (a) to continue and increase the trapped volume, by comparing with panel (b) we see that this added volume is expected to be unstable to small perturbations (in particular the trap's rf potential itself). At higher values of $U_{\rm rf}$ as in \fig{fig:Volume2D2}, restricting to lower values of $\lambda_{\rm b}$ is favorable.

The above simulations were based on constraining the maximal height in the calculation of the phase space volume as a function of the different trap parameters. Beyond being useful for numerical reasons, it is relevant experimentally for a few reasons.
Motion far from the trap center in any direction would be in the edges of the laser-cooling beam waist, and would damp slowly, possibly slower than typical heating rates. In particular, motion at high energy that approaches very closely the electrodes, is potentially more susceptible to heating effects which scale with $1/z^\alpha$, where $\alpha\approx 4$ \cite{brownnutt2015ion}. Finally, while being cooled towards lower energy, the ion could possibly hit the chaotic sea just because of the topology of the regular trapping islands, shown e.g.~in the higher energy shells of \fig{fig:PS2D2}.

We have truncated the simulations at the height given in \eq{eq:z0max}, which [with \eq{eq:physicalparams}] corresponds to a pseudopotential trap depth of at least $E/k_{\rm B}\approx 70\,$K [with $U_{\rm rf}=20\,$V and $U_{\rm b}=-0.2\,$V]. Since atomic ions are trapped in vacuum and routinely cooled by Doppler cooling to milliKelvin temperatures, with the entire setup possibly refrigerated to a few Kelvin, this is a very high energy scale. It is only smaller than the initial ion temperatures after their evaporation from an oven and ionization. We have not directly explored the motion at the scale of oven temperatures in our simulations, and the operation of laser cooling during the loading process remains an interesting open question.

\section{Experimental probe of non-equilibrium ion dynamics }\label{sec:experiment}

Experimentally probing the phase space structure of a trap is a challenging task. To avoid any non-Hamiltonian perturbations, the experiment has to be done on a short time scale (so that heating is negligible), and `in the dark' (with the laser cooling beam turned off after the initial cooling step). An ideal experimental sequence would consist of exciting the ion into various well-resolved initial conditions, and then measuring microscopic properties of the trajectory. Both aspects, however, are nontrivial to achieve with large amplitude motion in 3D. In this section we present an initial proposal of an experiment procedure that allows one to probe Hamiltonian transport in a mixed phase space with a trapped ion, while circumventing these difficulties. 
 
We consider an indirect method of exciting the ion to a large amplitude motion by an additional low amplitude periodic modulation \cite{wineland1983laser} of one electrode's potential (known as a `tickle'), that produces an oscillatory electric field at the equilibrium position of the ion. The tickle perturbation gradually (as a function of its amplitude) destroys the regular motion parts of phase space, turning them into mixed regions, with islands (that vanish and emerge), and chaotic strips that grow into contiguous chaotic regions. The latter may become connected to the regions already chaotic in the unperturbed Hamiltonian, with a structure and geometry which are very sensitive to the tickle parameters. Numerical simulations allow one to construct the ion's survival probability in the trap as a function of the tickle amplitude and its duration, which shows distinct features that can be averaged over the initial conditions. Thus this approach uses the properties of chaotic motion -- sensitive dependence on initial conditions, whose memory is however quickly erased (assuming it is mixing), to avoid the necessity of initializing the ion to precise initial conditions. It further replaces the difficulty of experimentally measuring the details of the trajectory with the more accessible reconstruction of the ion's survival probability as a function of the initial ion energy, the electric field amplitude of the tickle and the duration of the tickle. As we show below, the survival probability shows strong features that are distinguishable from simple thermal activation over a barrier, or a simple oscillator excitation process.
 
 \begin{figure} \hfill {\includegraphics[width=3.4in]{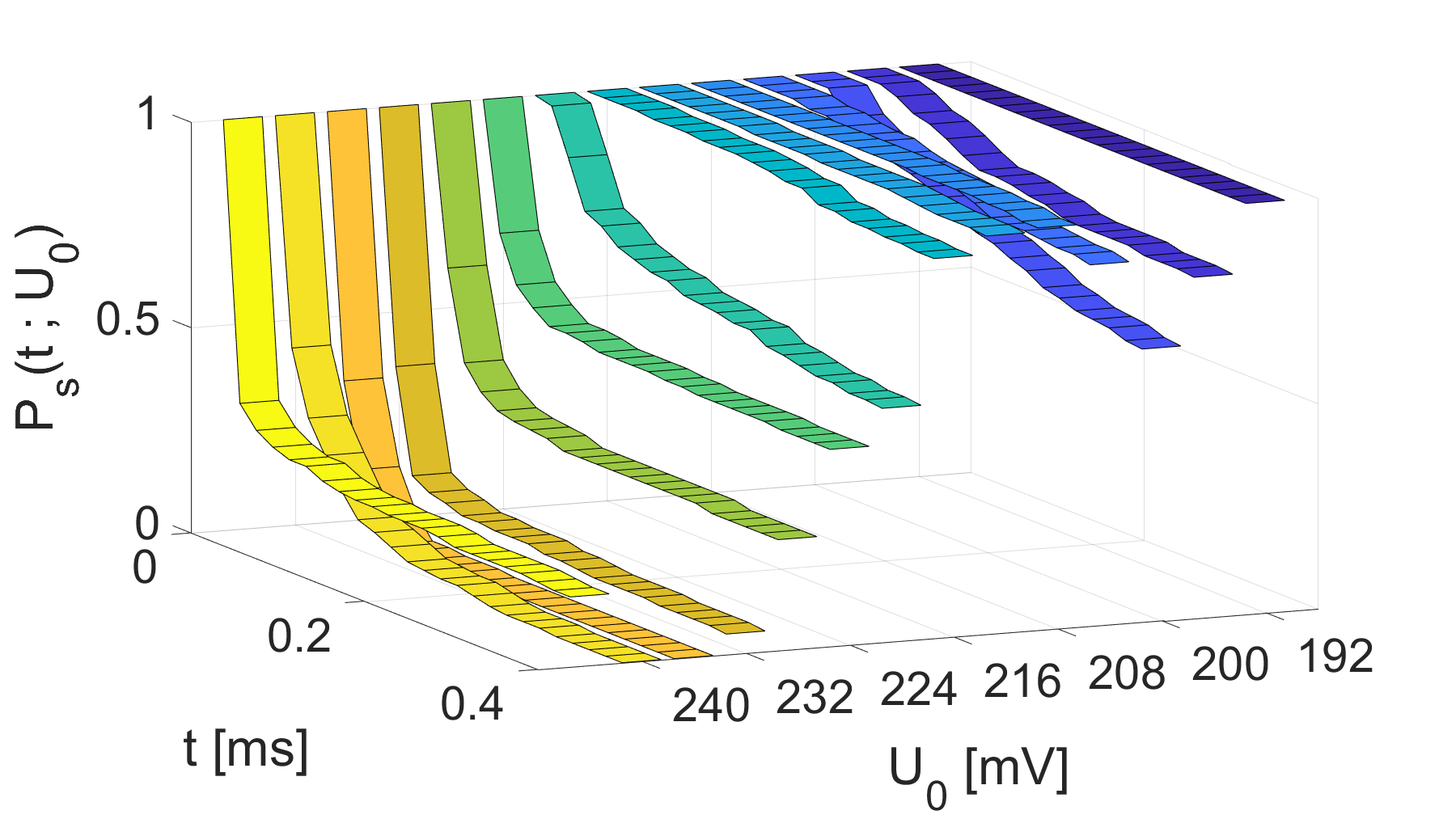} \hfill \caption{The simulated survival probability [\eq{Eq:Ps}] as a function of time for different tickle amplitudes ($U_0$), in a 4-wire trap (see the text for details of the model and parameters). The simulations show  a sharp threshold for the ion's escape out of the trap, followed by a non-monotonous dependence on $U_0$. We interpret the saturation of (some) curves as the co-existence of long-lived trapping pockets, that are destroyed and re-emerge as a function of the perturbation amplitude. These features together with the different functional forms of $P_s(t;U_0)$, present a potentially strong experimental signature that could be used to indirectly probe the trap's phase space structure, and a promising direction for studying Hamiltonian transport in a mixed phase space.} \label{fig:SurvivalP}} \end{figure}
 
We must now depart from the 5-wire trap in the symmetric configuration, modelled so far in this work. Hitherto, it was chosen because it enabled us to build the theory in steps of increasing complexity of the phase space and parameter dependence. In hindsight, it allowed us to uncover a near-integrable limit of surface traps, which plausibly stands at the basis of their practical success. However, in order to guarantee laser cooling in all three spatial directions, the $yz$ symmetry has to be broken in practice. This can be done in the 5-wire trap as discussed in the introduction, while another commonly used trap configuration is the so-called 4-wire trap.
The simulations described below are based on a 4-wire trap operational at NIST, which implies that the effects described could be observed in a realistic trap geometry. A detailed investigation of the phase space for this trap is beyond the scope of the current work. However, we model the experiment setup faithfully and by comparing the numerical simulations of the trap's rf potential to its pseudopotential approximation, we can also assess the correspondence between the two through their response to the tickle perturbation.

We model the trap used in \cite{wilson2014tunable}. The 4-wire trap potential can in general mix all three spatial coordinates (in the large amplitude regime), however for the  parameters of our simulation, we find that the axial motion (along $x$) remains to a large extent decoupled from the radial $yz$ plane. Nevertheless the $yz$ motion corresponds to a 5D phase space as discussed in \seq{sec:Hamiltonian2D}.
For the parameters of the simulation, the rf frequency is $\Omega=2\pi\times 163\,{\rm MHz}$, the stable fixed point is approximately $40\,\mu{\rm m}$ above the electrode surface and the linearized secular oscillation frequencies around it are 
\be \omega_{x,y,z}\approx 2\pi\times\left\{1.49,9.16,10.36\right\}{\rm MHz}.\label{Eq:Exp2}\ee
Numerical simulations are performed by solving the equations of motion of the ion in the 3D rf potential, starting with initial conditions picked at random from a thermal distribution at the Doppler cooling limit. A periodically modulated voltage (the tickle) is added to an  electrode potential \footnote{The tickle is applied in the simulation on electrode C1 in figure 1 of \cite{wilson2014tunable}. It is a very elongated electrode along the $x$-axis and consequently produces an electric field with a negligible component in this direction at the equilibrium position of the ion. The field components in the $y$ and $z$ directions are approximately equal and correspond to approximately 4 kV/m for a tickle amplitude of 1 V on the elongated electrode.}, with the form
\be U_{\rm d}(t)= {U}_{0}\cos({\omega}_{\rm d}t +\phi_0),\label{Eq:Udt1}\ee
where $\phi_0$, the initial phase of the tickle modulation, is a (uniformly distributed) random variable, and 
\be {\omega}_{\rm d}=1.1\times\omega_z.\label{Eq:baromegad}\ee
Simulating 200 initial conditions at each value of tickle modulation for a fixed maximal duration of $0.4\,$ms, we obtain the probability that the ion remains trapped as a function of time $t$ and tickle amplitude $U_0$. The ion's survival probability at fixed values of $U_0$ is shown in \fig{fig:SurvivalP}, for $U_0$ in the range $192-244\,$mV, in steps of $4\,$mV. We define the survival probability function  as
\be P_s(t;U_0) =1-P_e(\tau \le t;U_0),\label{Eq:Ps}\ee
where $P_e(\tau \le t;U_0)$ is the probability that the escape time $\tau$ (which is a random variable in each realization) is between 0 and $t$, for a fixed value of $U_0$.

With the frequency of the tickle modulation slightly higher than $\omega_z$, the center of the main island (close to resonant with the tickle), is gradually destroyed as $U_0$ is increased, but can remain as a pocket of long-lived chaotic motion (longer than the time reached in our simulation), protected from escaping by a broad region of (mostly) regular tori.
Indeed for tickle amplitude ${U}_{0}=192\,$mV we observe no escape of the ion. Increasing the tickle amplitude beyond this threshold value we find a sharp onset of ion escape. After a short-time flat part of the survival probability, the ion escapes at a slow rate, (almost linearly with time for $U_0=200\,$mV). Surprisingly, the escape rate is  not monotonic with the tickle amplitude, and $P_s(t;U_0)$ returns to nearly unity for $U_0=208-216\,$mV, before dropping sharply again. 

 \begin{figure} \hfill {\includegraphics[width=3.4in]{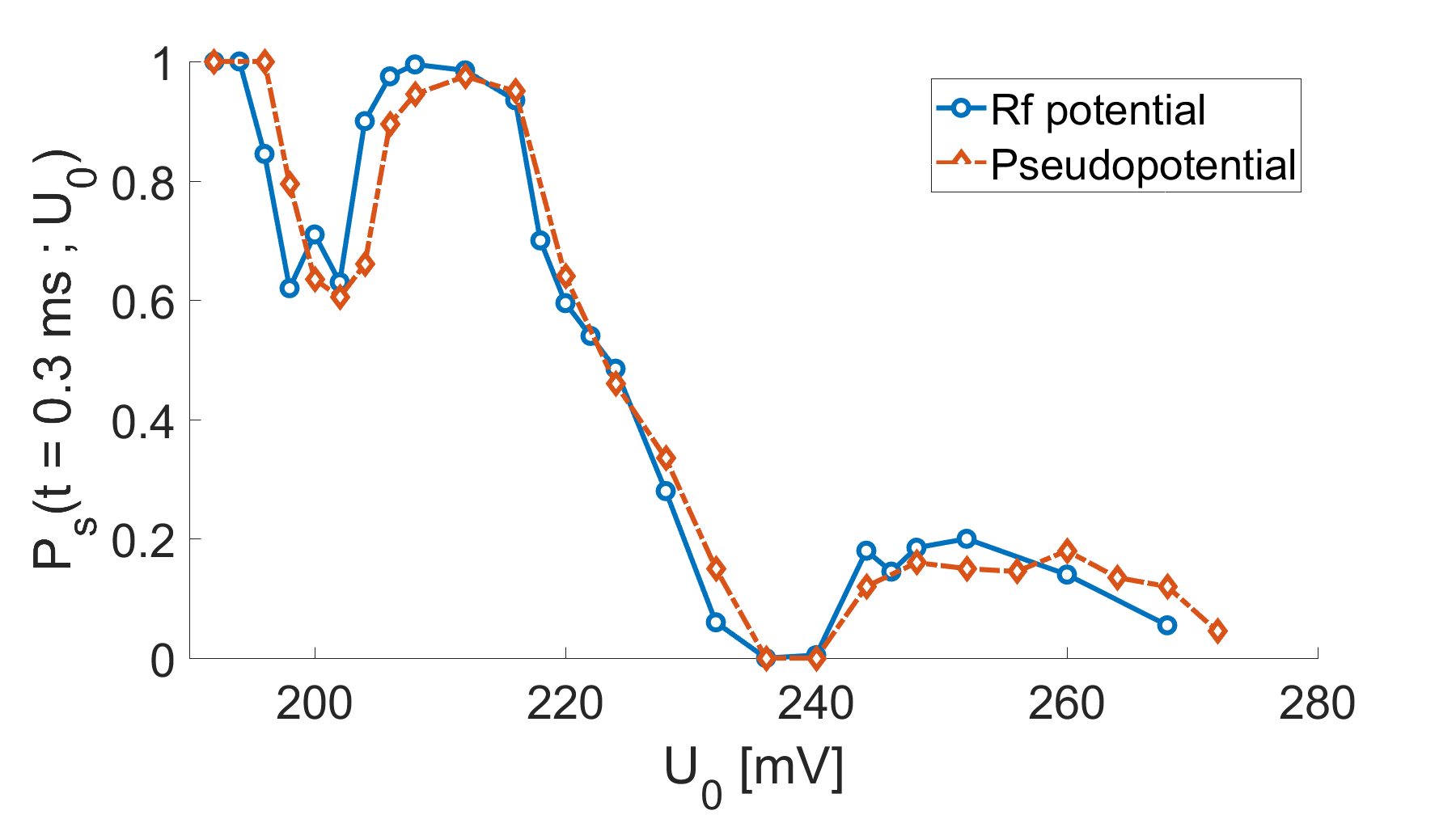} \hfill \caption{The survival probability at $t=0.3\,$ms as a function of the tickle amplitude, simulated using both the full rf potential of the trap and its pseudopotential approximation. The trap parameters are the same as in \fig{fig:SurvivalP}. The quantitative agreement of the two curves indicates that the simulated 4-wire trap is close to the limit where the rf potential is well described by its pseudopotential approximation.
 Finding such a
  non-monotonous dependence on $U_0$ in an experiment will distinguish the predictions of the phase space theory from a thermal or a parametric oscillator activation across a barrier.} \label{fig:SurvivalExp}} \end{figure}
 
With $U_0\approx 220\,$mV, the functional shape of $P_s(t;U_0)$ changes strongly and approximates an exponential distribution with a saturation. This is possibly the result of a branching process where the ion, depending on the exact (random) initial conditions, may get trapped in a long lived `pocket', or otherwise escape the trap. The  escape plausibly occurs from a (mixing) chaotic region through a `bottleneck' leading to the unstable fixed point, typically a memoryless process with an exponential distribution. The trapping is likely to occur in a large region in phase space that is connected to the region of initial conditions by chaotic motion, but weakly connected to the unstable fixed point, so the probability to escape the pocket is low on the studied timescale. On a longer time scale the trapping pockets could eventually decay. 
For $ {U}_{0}\approx 240\,$mV there are no observable pockets, but the ion lifetime increases with the perturbation amplitude, and new pockets are formed. 
 It should be stressed that the escape probability shown in \fig{fig:SurvivalP} is strictly different from the exponential decay of a simple thermal activation mechanism above a barrier, and also cannot be explained as a linear or parametric resonance of a harmonic oscillator, relevant in the low amplitude regime \cite{PhysRevA.94.023401,KinkTickle}. This rich nonmonotonic behaviour remains to be explored with a more detailed analysis of the phase space, the pockets and the statistical properties of the motion, but the salient features could be observed with a relatively simple experiment.

 Figure \ref{fig:SurvivalP} indicates that the escape
 probability beyond $t\gtrsim 0.1\,$ms is weakly dependent on 
 time. In \fig{fig:SurvivalExp} we plot the value of the survival
 probability at fixed $t=0.3\,$ms as a function of the tickle voltage
 (with data from more simulations). The non-monotonic dependence of
 the survival probability on $U_0$ is a clear experimental signal that
 may be obtained by just cooling the ion to the Doppler limit and
 varying the tickle amplitude.  We note that due to the signal being
 based on ensuing ion escape, a full set of experiments would require
 a large number of lost and reloaded ions. Beyond a certain number of
 repetitions this might degrade the trap stability due to uncontrolled
 stray charges of the trap electrodes produced by the loading, a
 technical weakness of the proposed scheme. An extension of the
 proposed experiment to a multi-zone trap, separating the loading
 region from the escape region, may avoid these shortcomings. Another
 possibility would be observing transfer between multiple wells
 without going over the trap's barrier, thus avoiding the ion's escape
 altogether.

Figure \ref{fig:SurvivalExp} also shows  the results of simulations using the pseudopotential approximation for the trap. We see that for these parameters the pseudopotential closely reproduces the effect of the tickle. This is an indirect indication that the trap is operating in a regime where its phase space structure is mostly determined by the pseudopotential with little perturbation from the time dependent drive. 
From the results of \seq{Sec:Hamiltonian1D} and \seq{sec:Hamiltonian2D}, we expect that this holds close to an integrable limit of the trap. In addition, we can conclude that the mechanisms leading to the rich transport features described above are occurring within a 5D phase space (that of the 4D pseudopotential with the additional time-dependent tickle), to which surface traps provide access for a controllable experimental study.
Further numerical studies of the trap's phase space and  complementing experiments would be required to confirm the validity of our findings, and to establish that the general conclusions of the theory presented in this work are relevant for a wide variety of surface electrode traps used in practice. 

\section{Summary and Outlook}\label{Sec:Outlook}

The main goal of this paper has been to provide a description of
Hamiltonian dynamics within surface electrode Paul traps. We have
aimed at developing a general framework that enables,
put simply, to understand both what is going on in the phase space,
and why it is like that. This  understanding of the nature of the
dynamics has furthermore allowed us to answer practical questions such
as  the optimal choice of experimental parameters to obtain
the largest possible trapping regions. 

To achieve that goal, we have increased the complexity of the treatment gradually, starting from a time independent one-dimensional model describing the motion along the $z$ direction perpendicular to the trap surface, and including successively the effects of the time dependence and of  coupling with the $y$ direction (in plane but orthogonal to the direction of the rf wires). In this process we find that although many regimes,
each of them having their own idiosyncrasy, exist, the global picture that emerges is
characterized by a few simple observations.  

The first  is that in the regime where presumably most traps are operated in practice, chaotic trajectories usually do not remain trapped. Thus we can identify the trapped volume with
the volume of regular (near-integrable) dynamics.  As a consequence, there is in general a
competition between increasing the strength of the applied trapping forces, which we would intuitively expect to increase trapped phase space
volumes, and the necessity to avoid entering a regime which is too chaotic.
The optimal choice of parameters is determined by that competition, controlled by a few parameters that play multiple roles, and for the 5-wire configuration that we have considered in detail,
we provide explicit values (while, of course, real life traps would always contain elements omitted from our model).

More broadly, the theory presented in this work gives a general framework that can be applied to various surface-electrode traps. It can be a starting point for asking more refined questions on the dynamics, for example including dissipative processes like cooling, or heating. Of the general conclusions that we can draw, an important first step in a future analysis would be to identify an integrable or near-integrable limit of a trap. Then, the parameters should be studied according to their role in increasing the phase space volume on the one hand, and introducing chaotic motion on the other hand. As we saw, typically each parameter introduces both effects, with some value forming a threshold between the two. A further general conclusion is that a classification of the phase space motion within the pseudopotential approximation could be sufficient for answering many questions, for practical parameters and regimes of motion. A natural and robust `boundary' is formed by those parameter values for which the motion is largely integrable within the pseudopotential, up to the point at which a change of the parameters introduces a lot of chaotic motion. The same holds for the rf potential, which has a threshold for the drive amplitude, beyond which a lot of chaos is quickly introduced. Thus below this threshold, for any new parameter that has to be considered, we expect that a study of the pseudopotential is sufficient (up to the point where a significant volume of chaotic motion is introduced), which implies a tremendous simplification. Above this rf threshold, at least for the examples analyzed here, we find that there is little benefit because the trajectories quickly escape.

 This work represents also a first and necessary step to study in detail Hamiltonian and non-Hamiltonian phase space transport effects. We have in this work merely touched on the complex motion possible in the 5D phase space. We have presented a preliminary study indicating that rich Hamiltonian transport mechanisms can be explored in the ion trap, and we hope to provoke further endeavours in this direction. The phase space framework is also a natural starting point for a semiclassical consideration of quantum mechanical effects.  
  The strength of a charged particle's interaction with the environment also implies that stringent conditions on the system's isolation are required in order to observe such effects. The manifestations of the phase space structures that we studied in this work, when the motion becomes sufficiently coherent \cite{bohigas1993}, could be an interesting future direction. 
 
Finally, a quantitative study of  non-Hamiltonian effects, such as heating by electric noise, or laser cooling dynamics within the Paul trap, is also of importance and interest. 
In particular, the combination of micromotion together with stochastic processes, presents a challenge both for experimental control and theoretical
description. 
We will study these effects in future publications. In \cite{rfcooling} we will study in detail laser cooling in the nonlinear and time-dependent potential, and its interplay with  position-dependent noise, where the phase space framework will again prove essential to our approach of the problem. Strong position-dependence is one characteristic of a dominant source of noise in Paul traps, the so-called anomalous heating \cite{brownnutt2015ion}, which is also characterized by a `colored' (frequency dependent) noise spectrum. The treatment of colored noise, again using a phase space approach, will be studied separately \cite{AnomalousNoise}.


\begin{acknowledgments}
We thank Ananyo Maitra for very helpful advice on the manuscript. H.L. thanks Giovanna Morigi, Alex Retzker and Roni Geffen for fruitful discussions.
H.L. acknowledges support by the French government via the 2013-2014 Chateaubriand fellowship of the French embassy in Israel, support by a Marie Curie Intra European Fellowship within the 7th European Community Framework Program, and support by IRS-IQUPS of Universit\'{e} Paris-Saclay.

\end{acknowledgments}

\begin{widetext}
\appendix

\section{The pseudopotential}\label{app:pseudopot}

The pseudopotential \cite{amini2008} is a time-independent approximation \cite{rahav2003effective} of the time-dependent rf potential. We will now derive the pseudopotential approximation using canonical perturbation theory \cite{cary1981lie}. We introduce a formal perturbation parameter $\epsilon$, and we consider a particle subject to the Hamiltonian 
\be H(q,p,t)=H_0(p)+\epsilon H_1(q,t)+\epsilon^2 H_2(q)=p^2 /2+\epsilon f(t)V_2(q)+\epsilon^2 V_0(q),\label{Eq:Hqptepsilon}\ee
where $f(t)$ is periodic with a period $T=2\pi/\Omega$. At the end of the calculation we can set $\epsilon=1$, but we will see that $\epsilon$ can be assigned a physical meaning, and that the ordering of the terms in \eq{Eq:Hqptepsilon} is meaningful as well. We avoid an explicit vector notation here, but the dimension of phase space is arbitrary. We are searching for a canonical transformation to the new set of conjugate coordinates $\{Q,P\}$, derived from a generating function $F(q,P,t)$. The transformation will depend explicitly on time, and is required to reduce to the identity at $t=0$. Starting with $q(t=0)$ and $p(t=0)$, we will construct the canonical transformation to describe the stroboscopic map of $H(q,p,t)$, such that $\{Q(t=T),P(t=T)\}$ approximate $\{q(t=T),p(t=T)\}$ within the Poincar\'{e} surface of section after one iteration. We will see that the evolution of  $\{Q,P\}$ can be derived from a time-independent Hamiltonian expanded in the perturbation parameter,
\be K(Q,P)=K_0(Q,P)+\epsilon K_1(Q,P)+\epsilon^2 K_2(Q,P)+...\label{Eq:Kexp}\ee
Hence with the canonical transformation defined by  $ q=q(Q,P,t)$ and  $ p=p(Q,P,t)$, we require that $F$ obeys
\be F(q,P,t)=qP+\epsilon F_1+ \epsilon^2 F_2+...\,, \qquad p = \frac{\partial F}{\partial q},\qquad Q = \frac{\partial F}{\partial P},\label{Eq:Fdef}\ee
and we expand similarly the transformation functions,
\be q(Q,P,t)=Q+\epsilon {q}_1+\epsilon^2 {q}_2+...\,,\qquad p(Q,P,t)=P+\epsilon {p}_1+\epsilon^2 {p}_2+...\label{Eq:qQPt}\ee
Now plugging $q=q(Q,P,t)$ using \eq{Eq:qQPt} into the definition of $F$ from \eq{Eq:Fdef}, Taylor expanding the terms of $F$ with respect to $q$, and collecting the terms up to order $\epsilon^2$, we have
\be p_1=\left[\frac{\partial F_1(q,P,t)}{\partial q}\right]_{q=Q},\qquad p_2=\left[\frac{\partial ^2 F_1(q,P,t)}{\partial q^2} q_1(Q,P,t)+ \frac{\partial F_2(q,P,t)}{\partial q} \right]_{q=Q},\label{Eq:Fpartials1}\ee
where the partial derivatives are to be taken with respect to argument $q$, and then $q$ is simply to be replaced by the variable $Q$ (and not by the function $q(Q,P,t)$, which has already been used in the expansion). Similarly,
\be q_1=-\left[\frac{\partial F_1(q,P,t)}{\partial P}\right]_{q=Q},\qquad q_2=-\left[\frac{\partial ^2 F_1(q,P,t)}{\partial q \partial P} q_1(Q,P,t)+ \frac{\partial F_2(q,P,t)}{\partial P} \right]_{q=Q}.\label{Eq:Fpartials2}\ee
Finally, the transformed Hamiltonian, expanded to the same order, is given by
\be K(Q,P)=H(q,p,t)+ \frac{\partial F(q,P,t)}{\partial t}=H_0(P+\epsilon p_1+\epsilon^2 p_2)+\epsilon H_1(Q+\epsilon q_1,t)+\epsilon^2 H_2(Q) + \epsilon\left[  \frac{\partial F_1}{\partial t}\right]_{q=Q+\epsilon q_1}+\epsilon^2\left[  \frac{\partial F_2}{\partial t}\right]_{q=Q},\label{Eq:Ktansfexp}\ee
where the first step of the expansion has been indicated, and the next steps will be performed in the following.

Equating \eq{Eq:Kexp} with \eq{Eq:Ktansfexp}, we get at order $\epsilon^0$ simply
\be K_0(P)=H_0(p=P)=P^2/2.\label{Eq:K_0}\ee 
We now recall that if $G(Q,P,t)$ is a function of phase space variables with an explicit dependence on time, its total time evolution (on an orbit in phase space, for motion generated by the Hamiltonian $K_0$) has the derivative (using Poisson brackets),
\be \frac{d G(Q(t),P(t),t)}{dt}=\frac{\partial  G(Q,P,t)}{\partial t}+\{K_0,G\}=\frac{\partial  G}{\partial t}+\frac{\partial  K_0}{\partial P}\frac{\partial  G}{\partial Q} -\frac{\partial  K_0}{\partial Q}\frac{\partial  G}{\partial P}=\frac{\partial  G}{\partial t}+P\frac{\partial  G}{\partial Q}.\label{Eq:Gt}\ee
At order $\epsilon^1$ we get the equation
\be K_1(Q,P)=p_1(Q,P,t)P + H_1(q=Q,t)+\left[\frac{\partial F_1}{\partial t}\right]_{q=Q},\label{Eq:F_1eq}\ee
which, substituting $p_1$ from \eq{Eq:Fpartials1} and using \eq{Eq:Gt}, can be seen to have the form
\be \frac{d G_1(Q,P,t)}{dt}=K_1(Q,P)-H_1(Q,t),\label{Eq:dGdt1}\ee
where $Q$ evolves along the orbits of $K_0(P)$ [a free particle], and $G_1=\left[ F_1\right]_{q=Q}$. Equation \eqref{Eq:dGdt1} can be solved by the method of integrating of the r.h.s along these orbits \cite{cary1981lie}, and $F_1$ would be that integral, with $q$ replacing $Q$ and the initial condition $F_1(t=0)=0$. Skipping the details of the derivation, it is easy to verify that $G_1$ is given by
\be G_1(Q,P,t)=\int_{0}^{t}\left[K_1(Q+P(t'-t),P)- H_1(Q+P(t'-t),t')\right]dt'.\label{Eq:G_1}\ee
and hence, $F_1(q,P,t)=\left.G_1(Q,P,t)\right|_{Q=q}$.

So far the treatment has been quite general, and there is still a lot of freedom in the choice of $K_1$. The natural choice that defines the pseudopotential, is to take $K_1$ to be the average of $H_1$ along the orbit of $K_0$, over a period $T$. This will make $\{Q,P\}$ the mapping of $\{q,p\}$ under one iteration of the stroboscopic map, and also $F_1$ will remain well-behaved -- without terms that are unbounded in time. We find
\be K_1(Q,P)=\frac{1}{ T}\int_0^T H_1(Q+P(t'-T),t')dt' \approx \frac{1}{T}\int_0^T f(t')[V_2(Q)+\nabla V_2(Q)P(t'-T)]dt', \label{Eq:K_1}\ee
where the potentials have been Taylor expanded to the required order (as will be justified below), and we use $\nabla =\partial / \partial Q$ to emphasize that the results can be generalized to any dimension as noted above. We will now further simplify the algebra in the following by assuming that $f(t)$ obeys the two conditions
\be \int_0^T f(t')dt' = 0, \qquad  \int_0^T f(t')t'dt' = 0,\label{Eq:ftconditions}\ee
which hold for any $T$-periodic and time-reversal-invariant function $f(t)=f(-t)$, and in particular for $f(t)=\cos\Omega t$. In this case, we get simply 
\be K_1(Q,P)=0.\label{Eq:K1}\ee We can now obtain the canonical transformation functions (at order $\epsilon$) explicitly,
\be p_1(Q,P,t)=\frac{\partial G_1}{\partial Q}= -\int_0^t dt'f(t')\nabla V_2(Q+P(t'-t)),\quad q_1(Q,P,t)=-\frac{\partial G_1}{\partial P}= \int_0^t dt'f(t')(t'-t)\nabla V_2(Q+P(t'-t)).\label{Eq:p1q1}\ee

At order $\epsilon^2$ we get the equation
\be K_2(Q,P)=p_2(Q,P,t)P + \frac{1}{2}\left[p_1(Q,P,t)\right]^2+\left[H_2(q)+\frac{\partial H_1(q,t)}{\partial q} q_1(Q,P,t)+\frac{\partial^2 F_1}{\partial q\partial t} q_1(Q,P,t) +\frac{\partial F_2}{\partial t} \right]_{q=Q},\label{Eq:F_2eq}\ee
which, as before, substituting $p_2$ from \eq{Eq:Fpartials1} and using \eq{Eq:Gt}, can be rearranged to have the form
\be \frac{d G_2(Q,P,t)}{dt}=K_2(Q,P)-H_2(Q) -\frac{1}{2}\left[p_1(Q,P,t)\right]^2 +\frac{\partial}{ \partial Q} \left[p_1(Q,P,t)P+H_1(Q,t)+\frac{\partial F_1(Q,P,t)}{ \partial t}\right]q_1(Q,P,t),\label{Eq:dG_2dt}\ee
with $G_2=\left[ F_2\right]_{q=Q}$.
From \eq{Eq:F_1eq}, the term in square brackets is just $K_1(Q,P)$ and by \eq{Eq:K1} it is 0.
 We will not need the explicit solution for $G_2$ [that can be obtained as in \eq{Eq:G_1}], however, we are interested in the pseudopotential $K_2(Q,P)$, that is chosen to make the mean of $G_2$ over a period $T$ vanish. Hence we have
\be K_2(Q,P)=\frac{1}{T}\int_0^T\left[H_2(Q+P(t'-T)) +\frac{1}{2}\left[p_1(Q+P(t'-T),P,t')\right]^2 \right]dt'.\ee
which gives, using \eq{Eq:p1q1},
\be K_2(Q,P)\approx V_0(Q)+ \frac{1}{2T}\int_0^Tdt'\left[\int_0^{t'}dt_2 f(t_2)\right]^2\left[\nabla V_2(Q)\right]^2.\ee

For $f(t)=\cos\Omega t$ we have thus derived the familiar pseudopotential expression (setting $\epsilon=1$),
\be K(Q,P)= P^2/2 + V_0(Q)+\frac{1}{4\Omega^2}\left[\nabla V_2(Q)\right]^2.\label{Eq:Kpseudo}\ee
We note that $K$ is independent of the momentum $P$ due to the conditions in \eq{Eq:ftconditions}.
We are left only with justifying the Taylor expansion of the potentials. The motion within the pseudopotential $K(Q,P)$ is characterized by a frequency $\omega$, and the non-dimensional frequency $\nu=2\omega/\Omega$ can serve as the perturbation parameter. The scale of the variables and functions above is given by
\be Q\sim 1,\qquad P\sim \omega Q\sim \frac{\nu}{\Omega}Q,\qquad K\sim V_0 \sim K_2\sim \omega^2 Q^2\sim {\nu^2}{\Omega^2}Q^2,\label{Eq:nuQorders} \ee
which is consistent with $V_2\sim\nu\Omega^2Q^2$ and  $\nabla V_2\sim\nu\Omega^2Q$ when \eq{Eq:Kpseudo} is used. Looking at the order of the terms in \eq{Eq:K_1} and \eq{Eq:dG_2dt} [for short times of order $\Omega^{-1}$] justifies a-posteriori the approximation in these equations, showing that pseudopotential of \eq{Eq:Kpseudo} is an approximation to second order in $\nu$. This is consistent with the perturbation parameter that was used in the derivation, $\epsilon=\nu$ \cite{rfions}, and the ordering of the terms in \eq{Eq:Hqptepsilon}.


\section{Embedding in 4D phase space}\label{App:Embedding4D}

In \eq{eq:HtoH} we have written the embedding of the motion generated by a general 1D time dependent Hamiltonian, in $H\pare{z,p_z,t} \rightarrow \mathcal{H}\pare{q,p,Q,P}=H\pare{q,p,Q}+P$.
To see that this embedding reproduces the original motion as solutions of Hamilton's equations, we can directly calculate the latter. The equations for $q$ and $p$ coincide with those of $z$ and $p_z$, and in addition we have $\dot{Q}=1$ (and hence $\dot{t}=1$ which is required by consistency), and $\dot{P}=-\partial H(q,p,Q) /\partial Q$. Since $\mathcal{H}$ is time-independent, each of its trajectories lies on a 3D manifold of constant `energy' $\mathcal{H}\pare{q,p,Q,P}\equiv\varepsilon$, in the 4D phase space. The original motion is reproduced for $\mathcal{H}=\varepsilon=0$, since this gives using \eq{eq:HtoH}, $0=H(t)+P(t)$, compatible with the equation for $\dot{P}$, as can be verified directly.

A Poincar\'e surface of section is constructed by taking simulations with different initial conditions at a fixed energy, and plotting a point in the plane of two variables (e.g.) $q$ and $p$, every time a simulated trajectory goes through a fixed value of (e.g.) $Q$, in a definite direction. Here since $\mathcal{H}$ depends on $Q$ only through $\cos 2Q$, we can understand that the stroboscopic map of $H$, taken at a fixed value of $t \pmod {\pi}$, is equivalent to a Poincar\'e surface of section of $\mathcal{H}$, taken within the fixed energy subspace $\varepsilon=0$, with $\cos 2Q={\rm const}$.  Therefore the phase space area within a closed curve of the stroboscopic map can be calculated using a method developed for calculating the action of 2D tori surfaces in the 4D phase space, described in \app{App:Action4D}. This method can also be applied to the calculation of the action for the $yz$ motion within the pseudopotential approximation, studied in \seq{sec:Hamiltonian2D}.

\section{Invariant tori in 4D phase space}\label{App:Action4D}

Here we describe (based on \cite{bohigas1993}) the calculation of action invariants in the 4D phase space of the canonical variables $\left\{q,p,Q,P\right\}$, that can correspond to the time-independent pseudopotential in 2 spatial dimensions, or to the time-dependent potential in 1 spatial dimension using the correspondence in \eq{eq:HtoH}.
On each invariant torus one can draw two independent paths $C_1$ and $C_2$, and define the corresponding area $J_a=\int_{C_a}\left[pdq+QdP\right]$, and action $I_a=J_a/2\pi$, with $a=1,2$. 
We are working within the integrable approximation (ignoring any KAM structure in the phase space), inside the main island up to the last unbroken torus, where there exists a local canonical action-angle transformation $\left\{q,p,Q,P\right\}\rightarrow (I_1,I_2,\theta_1,\theta_2)$.
The energy is conserved, i.e. 
\be \mathcal{H}(J_1,J_2)=\epsilon,\label{Eq:HJ1J2}\ee
and we can choose $C_1$ to be in the Poincar\'e surface of section defined by the intersection of the torus with $Q=0$. The action integral over $C_1$ is then $J_1=\int_{C_1}pdq$.
To obtain both the value of $J_1$ and $J_2$ (the latter cannot be calculated directly), we use the fact that \eq{Eq:HJ1J2} determines an implicit function $J_2(J_1)$. Taking its differential,
\be \frac{\partial \mathcal{H}}{\partial J_1}dJ_1 + \frac{\partial \mathcal{H}}{\partial J_2}dJ_2 = 0 \implies \frac{dJ_2}{dJ_1}=-\frac{\nu_1}{\nu_2}\equiv -\alpha,\ee
where $\alpha$ is the winding number of the torus and allows to define the Legendre transform of $J_2(J_1)$, 
\begin{align}
&J(\alpha)=J_2(J_1(\alpha))+\alpha J_1(\alpha).
\label{eq:J2alphaJ2}
\end{align}
The total action, $J$, is given by the integration of the Lagrangian over the time,
\begin{align}
J=\int\left[ {p}d{q}+PdQ\right]=\int  \left[ p\dot{q}+P\dot{Q}\right]dt.
\end{align}
The winding number can be approximated by $r$/$s$, where $r$ is the number of turns around the center of the Poincar\'e section and $s$ is the number of times, starting from $q\left(0\right),p\left(0\right)$ in the $Q=0$ plane, that the trajectory crosses the Poincar\'e section before returning (within some accuracy) to the initial point.
By the definition of the Legendre transform,
${dJ}/{d\alpha}=J_1$.
Therefore, simulating the equations of motion on a narrow grid  covering the phase space, and calculating the corresponding values of $J$ and $\alpha$, allows one to obtain $J_1$ and $J_2$. 

For the particular case of \eqs{Eq:H1Drf}-\eqref{eq:HtoH}, the action is given by
\begin{align}
J=\int  \left[ p\dot{q}+P\dot{Q}\right]dt
=\int\left[ p_z \dot{z}-H(z,p_z,t)\right] dt=\int\left[ \dot{z}^2-H(z,p_z,t)\right] dt= \int\left[ \frac{1}{2}\dot{z}^2-V_{\rm rf}^{\rm 1D}\pare{z,t}\right]dt.
\end{align}

\section{Volume in 4D phase space}\label{App:Volume4D}

To calculate any volume of a 4D phase space with a specific property, let us define an indicator function of phase space $\chi_{{\vec{q},\vec{p}\
}}$, taking the value 1 on phase space structures to be counted, and 0 otherwise. Here, $\vec{q}=\left\{q,Q\right\}$, and $\vec{p}=\left\{p,P\right\}$. Within a fixed energy shell $E$, the restricted 3D volume of interest can be defined by
 \be \Lambda^{\rm (3D)}_\chi(E)=\int { d}\vec{q}{ d}\vec{p} \left[\delta(E-H(\vec{q},\vec{p}))\chi(\vec{q},\vec{p})\right],\label{Eq:Vlambda3D2}\ee
 with the Hamiltonian $H(\vec{q},\vec{p})$.
We now employ a canonical transformation,
\be \{\vec{q},\vec{p}\}\to \vec{Z}\equiv \{q,p,t,-E\},\ee
where $\{q,p\}$ are the coordinates
 within the Poincar\'e section that has to be constructed. The integral in \eq{Eq:Vlambda3D2} then transforms according to
\be \Lambda^{\rm (3D)}_\chi=\int d\vec{Z}\left[\delta(E-H(\vec{Z}))\chi(\vec{Z})\right]=\\ \int dq dp T\left(q,p\right)\chi(q,p),\ee
where $T(q,p)$ is the return time of the particle to the Poincar\'e section - the time along the trajectory from a point starting on the Poincar\'e surface, to its next return to the section.
Then a total 4D phase space volume of interest up to energy $E$ can be obtained by integrating over the energy shells,
\be \Lambda^{\rm (4D)}_\chi(E)= \int_0^E  \Lambda^{\rm (3D)}_\chi(E'){ d}E'.\ee

\end{widetext}

\bibliographystyle{unsrt}
\bibliography{chaos_5wire}

\end{document}